\documentclass[amsmath,amssymb,twocolumn,aps,prd,nofootinbib,letterpaper,superscriptaddress,longbibliography,notitlepage]{revtex4-1}

\usepackage[usenames,dvipsnames]{color}

\usepackage{graphicx}

\usepackage{mathtools}
\usepackage{lmodern}

\usepackage{dblfloatfix}
\usepackage{csquotes}
\usepackage{amsmath}
\usepackage{amssymb}

\usepackage[caption=false]{subfig}
\usepackage{microtype}
\usepackage[english]{babel}
\usepackage[bookmarks]{hyperref}

\begin{document}

\title{Nonlinear/non-iterative treatment of EFT-motivated gravity}
\author{Ramiro Cayuso}
\email{rcayuso@perimeterinstitute.ca}
\affiliation{Perimeter Institute for Theoretical Physics, 31 Caroline Street North, Waterloo, Ontario, N2L 2Y5, Canada}
\affiliation{Department of Physics and Astronomy, University of Waterloo,
Waterloo, Ontario, Canada, N2L 3G1}
\author{Luis Lehner}
\email{llehner@perimeterinstitute.ca}
\affiliation{Perimeter Institute for Theoretical Physics, 31 Caroline Street North, Waterloo, Ontario, N2L 2Y5, Canada}

\date{\today}

\begin{abstract}
We study a higher derivative extension to General Relativity and present a
fully nonlinear/non-perturbative treatment to construct initial data and study its
dynamical behavior in spherical symmetry when coupled to a massless scalar field.
For initial data, we compare the obtained solutions with those from alternative treatments
that rely on a perturbative (or iterative) approach. For the future evolution of such data,
we implement a recently introduced approach which addresses mathematical pathologies
brought in by the higher derivatives. Our solutions demonstrate the presence of
unexpected phenomena---when seen from the lense of General Relativity, as well as departures from General Relativity in the quasi-normal mode behavior of the scalar field scattering off the black hole. 
\end{abstract}

\maketitle

\section{Introduction}

Understanding potential departures from General Relativity (GR) has long been a driving
goal in theoretical physics. With the availability of  data ---of increasing quality and quantity---spanning from cosmological to solar scales, intense efforts are being
invested to understand potential observations --or bounds-- on  ``beyond GR'' questions  
(e.g.~\cite{Will:2014bqa,Freire:2012mg,Baker:2012zs,Yunes_Siemens13,Berti_etal15,Yunes_etal16}).

A large body of theoretical efforts have furnished many interesting putative extensions to General
Relativity. Several of which have been scrutinized to different degrees especially in the context of 
cosmology (e.g.~\cite{Aghanim:2018eyx}),
solar system (e.g.~\cite{doi:10.1146/annurev.nucl.58.020807.111839}), 
binary pulsar timing (e.g.~\cite{Freire:2012mg}), near horizon-scale measurements around 
SgrA~\cite{2019A&A...625L..10G,2019ApJ...875L...6E}  and 
gravitational waves (e.g.~\cite{Abbott_etal16,PhysRevD.100.104036}). 
Naturally, the depth to which observations can 
probe different theories relies on the availability of specific predictions.  In the context of
extensions to GR, many such predictions
have been obtained within linearized regimes with respect to specific solutions (e.g. FRW in cosmology, and flat
spacetime)\footnote{
The implicit assumption in these efforts is that the particular theory under study, in the regime
considered, satisfies the property of {\em linearization stability} --the solution of the linearized problem
is representative of the solution to the non-linear problem in the linear regime--.}.

Incipient efforts are targeting nonlinear regimes, e.g. in binary pulsars, compact object coalescence, 
and black holes (e.g.~\cite{Barausse:2012da,Sagunski:2017nzb,Hirschmann:2017psw,Witek:2018dmd,Okounkova:2019dfo}). 
However, in such regimes, 
extracting predictions from most putative extensions to GR faces multiple 
formal and practical challenges. On the formal side, there is uncertainty on whether different theories
can define a well defined initial value problem in the regimes of
interest (nonlinear, strongly gravitating and possibly highly dynamical). At the practical level, the
desire to discern possible signatures with complex theories requires involved (and typically costly) 
numerical simulations, a problem that is exacerbated---and in many cases rendered formally impossible--- 
due to the mathematical challenge alluded above.

To elaborate further, we note such difficulties are typically quite serious when considering all but a few 
proposed extensions. Many present higher order derivatives, the possibility of
characteristics crossings and even a change in character of the underlying equations of motion 
(see e.g.~\cite{Ripley:2019hxt,Bernard:2019fjb}). The first can 
be responsible for a short-time blow up of
solutions, the second signals loss of uniqueness in the solution where crossing
takes place and the latter the demise of the initial value problem. Faced with these rather generic concerns, 
but determined to explore the theory,  one must find a way to explore the theory of interest for relevant
conditions. After all, it is in principle possible that 
within some neighborhood of relevant initial (and boundary) conditions, the formation of black hole formation hides problems in their
interiors, or dispersive behavior remedies high frequency problems while obstructions arise in non-relevant scenarios.
Nevertheless since numerical implementations of a 
given theory would most likely seed (by truncation/roundoff errors) components of the initial data outside possible safe neighborhoods, 
bad properties of the theory lurking in the shadows would render the simulation intractable.

To address the aforementioned difficulty, a few practical approaches have been 
proposed\footnote{For a simple example illustrating strengths and potential pitfalls see~\cite{Allwright:2018rut}.}:
(i) ``Reduction of order\footnote{Note this terminology is used in related but slightly different ways. Sometimes it
denotes order-reducing relevant equations (and in doing so replacing some problematic terms) 
and solving them iteratively/perturbatively. In others, it simply
refers to the latter, assuming corrections are small in the regime of study and evaluating problematic terms
in a passive way.}'' (e.g.~\cite{Witek:2018dmd,Okounkova:2019dfo}), 
(ii) ``Fixing the equations'' (see ~\cite{Cayuso:2017iqc}). 
The former adopts a perturbative approach
with respect to higher derivative corrections---which secularly modify the solution away from that of
GR---and the latter controls the higher frequency 
modes of the correction via an effective damping while incorporating the low modes from the get go.
As a prototypical example, consider viscous hydrodynamics. Approach (i) would incorporate viscous effects
secularly with respect to the solution obtained without it --since the zeroth order has viscosity off, the Reynolds'
number is infinite. Notably this would mean {\em all wavelengths} describing the
flow are subject to turbulence and the secular terms would have to ``fight this phenomena off'' for short wavelengths
Reynold's number to recover the laminar behavior. Approach (ii) on the other hand, would naturally 
accommodate both high and low Reynold's number regimes and dynamically damp short wavelength modes.
However, approach (ii) introduces an external adjustable parameter to execute such damping, and lack of sensitivity to such parameter
attest for the correctness of the solution. On the other hand, approach (i) would require working out at least a further
order to assess to what degree the solution can be trusted.

We have demonstrated the benefits of this approach in a few simplified model 
problems~\cite{Cayuso:2017iqc,Allwright:2018rut} and we now explore its application within the context
of a challenging extension to GR inspired from Effective Field Theory (EFT) considerations~\cite{Endlich:2017tqa} which also
explicitly unearths a number of delicate issues.
In the current work, adopting spherical symmetry for simplicity, we illustrate the application of the method
and address a number of required steps. In particular, we discuss the construction of initial data
consistent with the theory (and in passing also contrast with the reduction of order approach), the evolution of the system 
and impact of modifications to GR 
as well as relevant derivative operators required to discretize the higher derivatives.

This work is organized as follows. In Section~\ref{Model}, following~\cite{Endlich:2017tqa} we briefly discuss the theory adopted considering in our case also a minimally coupled self gravitating scalar field in the theory. We also
present the steps involved
for considering equation of motions governed by {\em second order} in time derivatives (but general spatial
derivatives). Section\ref{ID} describes the construction of initial data, deferring to section~\ref{results} the
results obtained and potential implications in section\ref{finalcom}. We included in the appendices further
information on numerical operators employed, and convergence results. Lastly, we employ geometrized units ($G=c=1$),
and use Greek or early Latin letters in the alphabet to denote spacetime indices and the latter part of the 
Latin alphabet for spatial indices.

\section{Model}\label{Model}
\subsection{EFT and field equations}
To fix ideas, and adopt a sufficiently challenging model, we here take an extension to GR constructed from an EFT point of view. In such approach, one introduces
no new degrees of freedom in the theory---as they are integrated out---and parameterizes new physics
through a suitable low-energy/long-distance expansion. New physics enters through local interactions organized
in terms of powers that depend on some given scale~\cite{burgess_2021}. Here, we consider the extension
presented in~\cite{Endlich:2017tqa} with the inclusion of a minimally coupled scalar field to endow the
target system (as we consider spherical symmetry) with non-trivial dynamics. We note that~\cite{Endlich:2017tqa}
builds the action for the EFT with the requirements that the theory respects unitarity, causality, locality, and includes
no new light degrees of freedom. These requirements are consistent with writing the most 
general Lagrangian by adding to the Einstein Hilbert action's terms that are constructed 
out of the Riemann tensor and suppressing them by a curvature scale comparable to the scale probed 
by gravitational wave observations. The action for this EFT reads,
\begin{equation} \label{Seff} 
S_{eff}= \int d^{4}x \sqrt{-g}2M_{pl}^{2}\left( R - \frac{\mathcal{C}^{2}}{\Lambda^{6}} -
\frac{\widetilde{\mathcal{C}}^{2}}{\widetilde{\Lambda}^{6}} -
\frac{\mathcal{C}\widetilde{\mathcal{C}}}{\Lambda_{-}^{6}} + \dots \right),
\end{equation}
where $\mathcal{C}\equiv R_{\alpha \beta \gamma \delta }R^{\alpha \beta \gamma \delta }$ and 
$\widetilde{\mathcal{C}}\equiv R_{\alpha \beta \gamma \delta }\widetilde{R}^{\alpha \beta \gamma \delta }$, 
with $\widetilde{R}^{\alpha \beta \gamma \delta } = \epsilon^{\alpha \beta}_{\quad \mu \nu } R^{ \mu \nu \gamma \delta }$, 
and the $+\dots$ correspond to sub-leading contributions.

Notice however, that the EFT built this way, starts with correction at $\Lambda^{-6}$ as it is restricted to
the vacuum. As argued in~\cite{deRham:2019ctd} in the non-vacuum case, interactions would
give rise to corrections at $\Lambda^{-4}$. More generally, depending on assumptions made, in principle other
orders could be present and a rigorous classification should be made to bring needed clarity in this discussion.
For concreteness however, we here stick to the model in~\cite{Endlich:2017tqa} so as to work in
a highly demanding (i.e. with respect to the order of derivatives to deal with) setting to stress our approach. 
To simplify somewhat the computational cost, we will also restrict to the case
 ${\widetilde{\Lambda}^{-6}}={\Lambda_{-}^{-6}}=0$)

We thus consider the action above and include a minimally coupled scalar field to obtain non-trivial
dynamics. The equations of motion are, 
\begin{align}
\begin{split}
G_{\mu \nu} & = 8\pi T_{\mu \nu} +\frac{1}{\Lambda^{6}}\Big( -8\,\mathcal{C}\, R_{\mu}^{\;\;\alpha}R_{\nu \alpha} + 8\,\mathcal{C}\,R^{\alpha \beta}R_{\mu \alpha \nu \beta}
\\ 
& + 4 \, \mathcal{C} \, R_{\mu}^{\;\;\alpha \beta \gamma}R_{\nu \alpha \beta \gamma} - \frac{1}{2} g_{\mu \nu}\mathcal{C}^{2} -4\mathcal{C}\nabla_{\mu}\nabla_{\nu}R
\\ 
& - 32R^{\beta \gamma \sigma \delta} \nabla_{(\mu}R_{\nu)}^{\;\; \alpha} \nabla_{\alpha}R_{\beta \gamma \sigma \delta} + 8\mathcal{C}\nabla_{\alpha}\nabla^{\alpha}R_{\mu \nu}
\\ 
& + 32R^{\beta \gamma \sigma \delta}\nabla_{\alpha}R_{\beta \gamma \sigma \delta}\nabla^{\alpha}R_{\mu \nu} + 8R_{\mu \;\; \nu}^{\;\; \alpha \;\; \beta} \nabla_{\beta}\nabla_{\alpha}\mathcal{C}\Big), \label{EEM}
\end{split}
\\[2ex]
&\nabla^{\mu}T_{\mu \nu} =0. \label{eqscalar_tuv}
\end{align}
with $G_{\mu \nu}$ the Einstein tensor and $T_{\mu \nu}$ the standard scalar field stress energy tensor with no potential.
Besides the presence of $T_{\mu \nu}$, the main difference between the field equations \eqref{EEM} 
with those in \cite{Endlich:2017tqa} is the appearance of terms with 
involving $R_{\mu\nu}$ and $R$, which vanish in their case.   
Clearly, modifications to GR in this theory are governed by involved higher-derivative/non-linear terms on the right-hand-side of Einstein's 
field equations. Demonstrating that our proposed method is capable of handling these equations is a central 
goal of this work.

\subsection{3+1 splitting}
We now discuss how we express our equations in a way amenable to numerical integration. To this end, we must face three
particular issues: (i) define a 3+1 initial value problem by a suitable spacetime decomposition, (ii) address the problem
of higher than second time derivatives in the resulting equations, (iii) address the related problem of higher order spatial derivatives
and the issue of well-posedness.

To start, we adopt the standard spacetime decomposition of spacetime in 3+1 form by introducing a spacelike foliation, with
intrinsic metric $\gamma_{ij}$, extrinsic curvature $K_{ij}$, and auxiliary lapse/shift variables $\{\alpha,\beta^i\}$. Further, we
adopt the (symmetric hyperbolic formulation, in the absence of corrections)  
 ``\textit{Generalized Harmonic}'' (GH) formulation of GR\cite{GH_old,Pretorius_2005,Lindblom_2006}. 
The full set of equations can be expressed as:
\begin{equation}\label{EEMuv}
 G_{\mu \nu} =   8\pi T_{\mu \nu} + \epsilon M_{\mu \nu},
\end{equation}
where we have also replaced ${\Lambda^{-6}}$ for $\epsilon$. 

Then the full system can be written as,
\begin{subequations}
\label{system_M}
\begin{align}
\partial_{\perp} \gamma_{i j}=&-2 \alpha K_{i j}, \label{subeq1}
\\
\begin{split}
\partial_{\perp} K_{i j}  = & \alpha\left[R^{(3)}_{i j}-2K_{i k} K_{j}^{k}-\widetilde{\pi} K_{i j}\right]-D_{i} D_{j} \alpha \\& -\alpha D_{(i} \mathcal{C}_{j)}-\kappa \alpha \gamma_{i j} \mathcal{C}_{T} / 2 \\ 
& -8 \pi G \alpha\left[S_{i j}-\gamma_{i j}(S-\rho) / 2\right] \\ & -\epsilon \alpha\left[S^{M}_{i j}-\gamma_{i j}(S^{M}-\rho^{M}) / 2\right], \label{subeq2}
\end{split}
\\
\partial_{\perp} \alpha =&\alpha^{2} \widetilde{\pi}-\alpha^{2} H_{T}, \label{subeq3}
\\
\partial_{t} \beta^{i} =&\beta^{j} \bar{D}_{j} \beta^{i}+\alpha^{2} \rho^{i}-\alpha D^{i} \alpha+\alpha^{2} H^{i}, \label{subeq4}
\\
\begin{split}
\partial_{\perp} \widetilde{\pi} =&-\alpha K_{i j} K^{i j}+D_{i} D^{i} \alpha+\mathcal{C}^{i} D_{i} \alpha -\kappa \alpha \mathcal{C}_{T} / 2 \\ & -4 \pi G \alpha(\rho+S) -\frac{\epsilon}{2} \alpha(\rho^{M}+S^{M}),\label{subeq5}
\end{split}
\\
\begin{split}
\partial_{\perp} \rho^{i}=& \gamma^{k \ell} \bar{D}_{k} \bar{D}_{\ell} \beta^{i}+\alpha D^{i} \widetilde{\pi}-\widetilde{\pi} D^{i} \alpha-2 K^{i j} D_{j} \alpha \\ & +2 \alpha K^{j k} \Delta \Gamma_{j k}^{i}+\kappa \alpha \mathcal{C}^{i}
\\
&-16 \pi G \alpha j^{i} -2\epsilon\alpha j_{M}^{i}, \label{subeq6}
\end{split}
\\
\nabla^{\mu}T_{\mu \nu}=&0,  \label{subeq7}
\end{align}
 \end{subequations}
with the constraints, 
\begin{subequations} 
\label{cons_M}
\begin{align}
\mathcal{C}_{T} & \equiv \widetilde{\pi}+K, \label{const1} \\
\mathcal{C}^{i} & \equiv-\rho^{i}+\Delta \Gamma_{j k}^{i} \gamma^{j k}, \label{const2} \\ 
\mathcal{H} & \equiv K^{2}-K_{i j} K^{i j}+R-16 \pi G \rho -2\epsilon \rho^{M}, \label{const3}\\ 
\mathcal{M}_{i} & \equiv D_{j} K_{i}^{j}-D_{i} K-8 \pi G j_{i} -\epsilon j_{i}^{M}, \label{const4}
\end{align}
\end{subequations}
where $K\equiv\gamma^{ij}K_{ij}$, ${D}_{i}$ and $\bar{D}_{i}$ are 
the covariant derivatives for the three-metric $\gamma_{ij}$ and the background 3-metric  $\bar{\gamma}_{ij}$ respectively. The derivative operator $\partial_{\perp}$ is 
defined as $\partial_{\perp}=\partial_{t} - \mathcal{L}_{\beta}$, where $ \mathcal{L}_{\beta}$ is the Lie derivative along the shift vector $\beta^{i}$. We define $\Delta 
\Gamma^{i}_{jk}:=^{(3)}\Gamma^{i}_{jk} - ^{(3)}\bar{\Gamma^{i}}_{jk}$ , where these are the Christoffel symbols for the induced metric and background metric (flat in spherical coordinates) respectively. We 
also define $H_{T}:= H^{\mu}n_{\mu}$, where $n_{a}$ is the normal vector to the spatial hypersurfaces defined by the spacetime foliation
(note, for completeness sake we include the gauge source vector $H_{\mu}$ for reference but in our studies it was sufficient to adopt $H_{\mu}=0$).
 We also introduce new dynamical variables 
$\widetilde{\pi}$ and $\rho^{i}$  through equations (\ref{subeq3}-\ref{subeq4}) to make the system (ignoring the extensions to gravity) first order in time derivatives. $S_{ij}$, $S$, $\rho$ 
and $j^{i}$ are the matter variables constructed from the Energy-Momentum tensor $T_{\mu\nu}$ as, $S_{ij}= P^{\mu}_{i}P^{\nu}_{j} T_{\mu\nu}$, its trace $S = \gamma^{ij}S_{ij} $,  
$\rho= n_{\mu}n_{\nu}T^{\mu \nu}$, and $j^{i}=-P^{i\mu}n^{\nu}T_{\mu\nu}$. Here the definitions for $S^{M}_{ij}$,$S^{M}$, $\rho^{M}$ and $j_{M}^{i}$  are analogous to the ones for the 
matter sources, but instead of using $T_{\mu\nu}$ we use $M_{\mu\nu}$. In addition, we now have also included now equation \eqref{subeq7} that determines the evolution for the matter degrees of freedom. \\

Let us now analyze the nature of the additional terms $M_{\mu\nu}$ we have incorporated into Einstein's equations. All these terms contain nonlinear combinations of derivatives of metric components with a combined scaling of $\lambda^{-8}$ (with $\lambda$ the local wavelength).
In particular, terms contain derivatives of order as high as fourth. Thus, such terms are present 
for the effective sources $S^{M}_{ij}$, $S^{M}$, $\rho^{M}$ and $j^{M}_{i}$\eqref{system_M}. Well-posedness has now clearly gone out the window. Both the presence of high order time derivatives---which bring 
forth so called Ostrogradsky's instability\cite{Ostrogradsky:1850fid}--- as well as higher order spatial derivatives (of both even and odd orders) doom 
prospects of defining well-posed problems for general cases. However, restriction of the initial data considered and control 
of potential pathologies introduced 
might enable obtaining well-posedness. In what follows we describe how these issues are addressed.

\subsection{Time derivative order reduction of the modified equations}\label{time_reduct}
We now turn our attention now to dealing with higher than second order time derivatives. To
do so, we follow a field redefinition approach (e.g.~\cite{Solomon:2017nlh}) whereby
higher time derivatives are expressed in terms of spatial derivatives by repeated use of
the field equations.

For presentation clarity, we illustrate this approach schematically and ignoring contributions from the matter terms.
First we rewrite system \eqref{system_M} in terms of variables $g^{a}=\lbrace\gamma_{ij},\alpha,\beta\rbrace$ by means of equations \eqref{subeq1},\eqref{subeq3}, and \eqref{subeq4}. Then equations \eqref{subeq2}, \eqref{subeq5} and \eqref{subeq6} can be cast as,
\begin{equation}\label{eq_eps_sqr}
\begin{split}
    \frac{\partial^{2} g^{a}}{\partial t^{2}} &=  \Delta^{a}(g,\partial_{\mu}g,\partial^{2}_{i}g) \\&+ \epsilon M^{a}(g,\partial_{\mu}g,\partial^{2}_{\mu}g,\partial^{3}_{\mu}g,\partial^{4}_{\mu}g) + \mathcal{O}(\epsilon^{2}),
\end{split}
\end{equation}
where $\Delta^{a}$ represents the contributions of GR that, as $(g,\partial_{\mu}g,\partial^{2}_{i}g)$ indicates, depend only on the variables $g^{a}$, their first spacetime derivatives, and their second spatial 
derivatives. The symbol $M^{a}$ encodes the contributions of extensions to GR's equations which depend on the variables $g^{a}$, their first, second, third and fourth spacetime derivatives.  Now, take equations \eqref{eq_eps_sqr} to  $\mathcal{O}(\epsilon)$, 
\begin{equation}\label{eq_eps}
    \frac{\partial^{2} g^{a}}{\partial t^{2}} =\Delta^{a}(g,\partial_{\mu}g,\partial^{2}_{i}g) + \mathcal{O}(\epsilon),
\end{equation}
and define higher time derivatives of the $g^{a}$ variables by suitable derivatives of \eqref{eq_eps}. For instance, the third time derivative, would be given to this order by,
\begin{equation}\label{third}
    \frac{\partial^{3} g^{a}}{\partial t^{3}} =\frac{\partial_{t}\Delta^{a}}{\partial t}(g,\partial_{\mu}g,\partial^{2}_{\mu}g,\partial_{\mu}\partial^{2}_{i}g) + \mathcal{O}(\epsilon).
\end{equation}
Notice the right-hand-side of \eqref{third} depends on second time derivatives of the $g^{a}$ variables, which can be re-expressed through \eqref{eq_eps}. This procedure can be repeated to express all higher than second order time derivatives appearing on the right-hand-side in terms of spatial derivatives 
(of high order) while keeping time derivatives to at most first order.

Armed with these definitions, and replacing in $M_{\mu \nu} \rightarrow \widetilde{M}_{\mu \nu}$, one
has 
\begin{equation}\label{M_rep}
 \widetilde{M}_{\mu \nu} = M_{\mu \nu} + \mathcal{O}(\epsilon),
\end{equation}
and can re-express the $M^{a}$ terms in equations \eqref{eq_eps_sqr} to yield the ``time reduced'' evolution equations which are unchanged to $\mathcal{O}(\epsilon)$,
\begin{equation}\label{eq_eps_sqr_rep}
\begin{split}
    \frac{\partial^{2} g^{a}}{\partial t^{2}} & =\Delta^{a}(g,\partial_{\mu}g,\partial^{2}_{i}g) \\&+ \epsilon \widetilde{M}^{a}(g,\partial_{u}g,\partial_{u}\partial_{i}g,\partial_{u}\partial^{2}_{i}g,\partial_{u}\partial^{3}_{i}g) + \mathcal{O}(\epsilon^{2}),
\end{split}
\end{equation}
Finally, we reintroduce variables $u^{a}=\lbrace K_{ij},\widetilde{\pi},\rho^{i}\rbrace$
(through \eqref{subeq1},\eqref{subeq3}, and \eqref{subeq4}) to present the system in
a first order in time form for the whole set of variables $v^{a}= \lbrace g^{a},u^{a}\rbrace$,
\begin{equation}\label{eq_rep_first_order}
\begin{split}
    \frac{\partial u^{a}}{\partial t} &=\Delta^{a}(v,\partial_{i}v,\partial^{2}_{i}g) \\&+ \epsilon \widetilde{M}^{a}(v,\partial_{i}v,\partial^{2}_{i}v,\partial^{3}_{i}v,\partial^{4}_{i}g) + \mathcal{O}(\epsilon^{2}).
\end{split}
\end{equation}
Equations \eqref{system_M} and \eqref{cons_M} are modified solely by replacing $S^{M}_{ij}$, $S^{M}$, $\rho^{M}$ and $j_{M}^{i}$ by $S^{\widetilde{M}}_{ij}$, $S^{\widetilde{M}}$, $\rho^{\widetilde{M}}$ 
and $j_{\widetilde{M}}^{i}$, constructed using the tensor $\widetilde{M}_{\mu \nu}$ instead of the tensor $M_{\mu \nu}$.
The only time derivatives are on the left-hand side of the equations, while the right-hand side has up to third order spatial 
derivatives for the  $\lbrace u^{a}\rbrace$ variables, and up to fourth order spatial derivatives for the $\lbrace g^{a}\rbrace$ variables. 
Later on, we will use a similar strategy to deal with the constraint equations when constructing consistent initial data.

Before moving on, we note that in the case of spherical symmetry, there is yet another step we
can take. One can make use of the constraint equations 
(and their spatial derivatives) \eqref{const3} and \eqref{const4} to replace high-order spatial derivatives of metric 
variables in $M_{\mu \nu}$ in terms of (higher) derivatives of the scalar field. For convenience, we do so here and, as 
a result, our equations of motion will not display derivatives of higher than second order in the metric. Instead,
there will be non-linear combinations of derivatives up to order two in the metric and higher derivatives of the scalar
field.

\subsection{Dealing with higher spatial derivatives. ``Fixing the equation''}
Having removed all higher order time derivatives we are not done as even without potential Ostrogradsky instabilities there is a long road ahead to ensure the well-posedness of an initial value problem. 
The existence of higher order derivatives (in this case of the scalar field), and non-linear terms describing 
products of up to second order spatial derivatives (of the metric variables) are responsible for a variety of problems
preventing the definition of a well-posed problem (at the analytical, and therefore numerical levels).
For instance, in \cite{Cayuso:2017iqc} several examples of simple toy models illustrate the problematic behavior that higher
derivatives can bring. Clearly, a suitable approach must be devised to even aspire to explore the theory of interest.
As mentioned, at present two options are being explored to address this issue: (i) ``\textit{Reduction of order}'' procedure\footnote{While related, this is not be be mistaken with the time reduction of 
order used in the  previous section.} and (ii) ``\textit{Fixing the Equations}''. In this work we choose the latter approach. We will 
devote this section to giving details on the implementation of this technique. Further details and 
motivations for this approach can be found in~\cite{Cayuso:2017iqc,Allwright:2018rut}. At its
core, such approach introduces an evolution prescription to the higher terms to ensure high frequency modes
are controlled. 

To this end, we introduce a new dynamical tensor $\Pi_{\mu \nu}$ with an evolution prescription
to dynamically constrain it to $\widetilde{M}_{\mu \nu}$ (and with initial data $\Pi_{\mu \nu}=\widetilde{M}_{\mu \nu}$). We write system \eqref{eq_rep_first_order} (omitting the $\mathcal{O}(\epsilon^{2})$ symbol) as,
\begin{align}
    \frac{\partial u^{a}}{\partial t} &=\widetilde{\Delta}^{a}(v,\partial_{i}v,\partial^{2}_{i}g) + \epsilon \Pi^{a},   \label{eq_rep_first_order_u}\\
    \tau\frac{\partial \Pi_{\mu \nu}}{\partial t } &= -(\Pi_{\mu \nu}-\widetilde{M}_{\mu \nu}(v,\partial_{i}v,\partial^{2}_{i}v,\partial^{3}_{i}v,\partial^{4}_{i}g)), \label{eq_rep_first_order_Pi}
\end{align}
where now $\Pi^{a}$ is computed using the tensor $\Pi_{\mu \nu}$ instead of the tensor $\widetilde{M}_{\mu\nu}$. Equations \eqref{eq_rep_first_order_Pi} are ad-hoc equations introduced to control $\Pi_{\mu\nu}$ to approach $\widetilde{M}_{\mu \nu}$ in 
a timescale given by the free parameter $\tau$. (Note, $\tau$ has dimensions of time; throughout this work specific values
will be given to it with respect to the total mass $M$ of scenarios considered.) 
The particular form of equation \eqref{eq_rep_first_order_Pi} is not unique 
though it should not be crucial as long as the solution remains well-behaved and within
the domain of applicability of the EFT. 
In such scenario, the physics obtained would be independent of the choice of equation, as  
well as the value of the damping timescale $\tau$. This approach 
controls the behavior of short wavelength modes in the original equations while preserving the 
physics at the long wave-length regime.

We thus arrive to the final form of the equations that are now ready for numerical 
implementation. Notice that in equations  \eqref{system_M} and \eqref{cons_M} one replaces 
 $S^{M}_{ij}$,$S^{M}$, $\rho^{M}$ and $j_{M}^{i}$ by $S^{\Pi}_{ij}$,$S^{\Pi}$, 
 $\rho^{\Pi}$ and $j_{\Pi}^{i}$, which are constructed with  
 $\Pi_{\mu \nu}$ instead $M_{\mu \nu}$. Additionally, one  
 incorporates equations \eqref{eq_rep_first_order_Pi} for the evolution of the new dynamical 
variables $\Pi_{\mu \nu}$.

\section{Target problem}
We study a simple case that is dynamic and in which non-linearities become relevant. To this end we consider the 
dynamics of a  spherically symmetric spacetime minimally coupled 
to a scalar field that induces non-trivial dynamics in the problem. 
The line element for our problem is,
\begin{equation} \label{ds}
\begin{split}
ds^{2}&=(-\alpha^{2}+g_{rr}\beta^{2})dt^{2} +2\beta g_{rr}drdt + g_{rr}dr^{2} \\ &
+ r^{2}g_{T}(d\theta^{2} + \sin^{2}\theta d\varphi^{2}).
\end{split}
\end{equation}
In these coordinates the general form of the tensor $\widetilde{M}_{\mu \nu }$ encoding 
the extension to GR takes the form, 
\begin{equation}\label{Mmunu}
    \widetilde{M}_{\mu \nu} = \begin{pmatrix} 	\widetilde{M}_{tt} & \widetilde{M}_{tr} & 0 & 0\\ \widetilde{M}_{tr} &\widetilde{M}_{rr}& 0 &0 \\ 0 & 0 &\widetilde{M}_{T}& 0 \\ 0 & 0 & 0 &\widetilde{M}_{T} \sin^{2}\theta  \end{pmatrix},
\end{equation}
with four independent components. The structure $\Pi_{\mu \nu} $ is also of the form of \eqref{Mmunu}.

The equation of motion for the massless scalar field \eqref{eqscalar_tuv} is,
\begin{equation}\label{scalar_field}
    \nabla^{\mu}\nabla_{\mu}\phi=0, 
\end{equation}
and we introduce the new variable $\Sigma$ defined by, 
\begin{equation}
    \Sigma := \frac{1}{\alpha}(\beta \partial_{r}\phi - \partial_{t}\phi) , \label{eq_rep_first_order_sigma}
\end{equation}
to also express the scalar field evolution equations in terms of first order in time derivatives.
 
\section{Initial Data \& implementation}\label{ID}
We next discuss how to construct initial data that 
is consistent with the modified theory we are working with. The procedure is similar to 
the one usually followed in GR, but there are certain unique aspects 
to be treated carefully.

We start with the usual conformal decomposition of the spatial metric,
\begin{equation}
    \gamma_{ij} = \psi^{4}\overline{\gamma}_{ij},
\end{equation}
where $\psi$ is the conformal factor and $\overline{\gamma}_{ij}$ is a given background metric which 
we take to be the flat metric in spherical coordinates. With this choice the extension to 
the \textit{Hamiltonian Constraint} takes the form,
\begin{equation}
 8 \nabla^{2}_{\textit{flat}}\psi + \psi^{5}(A_{ij}A^{ij}- \frac{2}{3}K^{2}) + 16\pi\psi^{5}\rho + 2\epsilon\psi^{5}\rho^{\widetilde{M}}=0, 
\end{equation}
where $A_{ij}$ is the traceless part of the extrinsic curvature tensor $K_{ij}$ and now the additional term $ 2\psi^{5}\rho^{\widetilde{M}}$ contains the modifications to GR. Notice that 
the effective energy density defined by extension to GR is now directly constructed using the time reduced 
tensor $\widetilde{M}_{\mu \nu}$. 

The extension to the \textit{Momentum Constraint} takes the form,
\begin{equation}
 \nabla_{j}A^{ij} - \frac{2}{3}\nabla^{i}K - 8\pi j^{i} - \epsilon j^{i}_{\widetilde{M}}=0,
\end{equation}
which includes the additional current-like term $- \epsilon j^{i}_{\widetilde{M}}$. 
We aim to define initial data with  
traceless extrinsic curvature $K_{ij}$ (i.e. $K=0$), so we adopt the following ansatz for 
$A_{ij}$,
\begin{equation}
    A_{ij} = \begin{pmatrix} 	K_{rr} & 0 & 0 \\ 0 &-r^2\frac{K_{rr}}{2} & 0 \\ 0 & 0 &-r^2\frac{K_{rr} \sin^{2}\theta}{2}  \end{pmatrix}.
\end{equation}
The resulting (extended) \textit{Hamiltonian} and \textit{Momentum} constraints are,
\begin{equation}\label{ini_eq_psi}
\begin{split}
 \frac{\partial^{2} \psi}{\partial r^{2}} & = -\frac{2}{r}\frac{\partial\psi}{\partial r}- \frac{3}{16}\frac{K_{rr}^{2}}{\psi^{3}}  - \pi \psi \left(\frac{\partial \phi}{\partial r}\right)^{2} 
 -\pi\psi^{5}\Sigma^{2} \\ 
 & + \epsilon\frac{\psi^{5}}{\alpha^{2}}\left(- \frac{\beta^{2}}{4}\widetilde{M}_{rr}  + \frac{\beta}{2}\widetilde{M}_{tr} - \frac{1}{4}\widetilde{M}_{tt}\right), 
\end{split} 
\end{equation}
\begin{equation}\label{ini_eq_Krr}
\begin{split}
 \frac{\partial K_{rr}}{\partial r} &= -2\psi^{-1}K_{rr}\frac{\partial \psi}{\partial r} - \frac{3}{r}K_{rr} 
 + 8\pi \psi^{4}\Sigma\frac{\partial \phi}{\partial r } \\
 & + \epsilon \frac{\psi^{4}}{\alpha}\left( \beta\widetilde{M}_{rr} - \widetilde{M}_{tr}\right),
\end{split} 
\end{equation}
respectively. 
For $\epsilon=0$ these are familiar forms in  GR, and
given appropriate boundary values, and initial data for the scalar field, a unique solution 
can be found. Notice, these equations do not depend on the gauge variables $\{\alpha,\beta^i\}$.
However, 
when $\epsilon\neq0$ the modifications to gravity add terms with high order spatial derivatives, 
highly nonlinear terms, and even a dependency on the gauge variables. 
To date, a thorough mathematical analysis for these types of equation in general cases is still lacking. Notice,
in particular, the presence of higher derivatives
require additional boundary conditions---either explicitly or implicitly given. We have explored two
ways of constructing initial data consistent  with this system. The first one involves a procedure similar to the one we used for the time 
order reduction in section \ref{time_reduct}---so as to express higher derivatives in terms of lower ones---and a second one which
is essentially the iterative approach (i) mentioned in the introduction. For clarity we will refer to them as {\em order 
reduced} and {\em iterative} methods.  

\subsection{Order-reduced direct integration (ORDI)}
In this approach one replaces high-order spatial derivatives on the tensor $\widetilde{M}_{\mu \nu}$
by means of equations \eqref{ini_eq_psi} and \eqref{ini_eq_Krr} to the order desired. 
Schematically, to first order in $\epsilon$, we can write equations  \eqref{ini_eq_psi} and \eqref{ini_eq_Krr} as,
\begin{equation}\label{ini_eq_psi_schem}
 \frac{\partial^{2} \psi}{\partial r^{2}}  = \Psi(\chi,\partial_{r}\psi) + \epsilon\widetilde{M}_{\psi}(\chi,\partial_{r}\chi,\partial^{2}_r\chi,\partial^{3}_{r}\chi,\partial^{4}_r\psi),
\end{equation}
\begin{equation}\label{ini_eq_Krr_schem}
 \frac{\partial K_{rr}}{\partial r} = \mathcal{K}(\chi,\partial_{r}\psi) + \epsilon\widetilde{M}_{K}(\chi,\partial_{r}\chi,\partial^{2}_r\chi,\partial^{3}_{r}\chi,\partial^{4}_r\psi), 
\end{equation}
were $\chi=\lbrace \psi,K_{rr} \rbrace$ and we omit (in the presentation) the 
matter variables since their initial values are chosen freely. By neglecting  
$ \mathcal{O}(\epsilon)$ terms in equations \eqref{ini_eq_psi_schem} and \eqref{ini_eq_Krr_schem} then the expressions,
\begin{align}\label{ini_reduc}
  \frac{\partial^{2} \psi}{\partial r^{2}} & = \Psi(\psi,\partial_{r}\psi,K_{rr}) +  \mathcal{O}(\epsilon), \nonumber \\
  \frac{\partial K_{rr}}{\partial r} & = \mathcal{K}(\psi,\partial_{r}\psi,K_{rr})  +  \mathcal{O}(\epsilon),
\end{align}
\begin{align}\label{ini_3_dev}
  \frac{\partial^{3} \psi}{\partial r^{3}} & = \frac{\partial \Psi}{\partial r}(\chi,\partial_{r}\chi,\partial^{2}_{r}\psi) +  \mathcal{O}(\epsilon), \nonumber \\
  \frac{\partial^{2} K_{rr}}{\partial r^{2}} & = \frac{\partial{\mathcal{K}}}{{\partial r}}(\chi,\partial_{r}\chi,\partial^{2}_{r}\psi)  +  \mathcal{O}(\epsilon),
\end{align}
\begin{align}\label{ini_4_dev}
  \frac{\partial^{4} \psi}{\partial r^{4}} & = \frac{\partial^{2} \Psi}{\partial r^{2}}(\chi,\partial_{r}\chi,\partial^{2}_{r}\chi,\partial^{3}_{r}\psi) +  \mathcal{O}(\epsilon), \nonumber \\
  \frac{\partial^{3} K_{rr}}{\partial r^{3}} & = \frac{\partial^{2}{\mathcal{K}}}{{\partial r^{2}}}(\chi,\partial_{r}\chi,\partial^{2}_{r}\chi,\partial^{3}_{r}\psi)  +  \mathcal{O}(\epsilon),
\end{align}
can be used to redefine $\widetilde{M}_{\psi}$ and $\widetilde{M}_{\psi}$ in terms of only  $\lbrace \psi,\partial_{r}\psi,K_{rr} \rbrace $, so that,
\begin{align}
 \overline{M}_{\psi}(\psi,\partial_{r}\psi,K_{rr}) & = \widetilde{M}_{\psi}(\chi,\partial_{r}\chi,\partial^{2}_r\chi,\partial^{3}_{r}\chi,\partial^{4}_r\psi) + \mathcal{O}(\epsilon),  \nonumber \\
 \overline{M}_{K}(\psi,\partial_{r}\psi,K_{rr}) & = \widetilde{M}_{K}(\chi,\partial_{r}\chi,\partial^{2}_r\chi,\partial^{3}_{r}\chi,\partial^{4}_r\psi) + \mathcal{O}(\epsilon).
\end{align}
Finally the system of equations \eqref{ini_eq_psi_schem} and \eqref{ini_eq_Krr_schem} can be redefined as,
\begin{equation}\label{ini_eq_psi_schem_R}
 \frac{\partial^{2} \psi}{\partial r^{2}}  = \Psi(\psi,\partial_{r}\psi,K_{rr}) + \epsilon\overline{M}_{\psi}(\psi,\partial_{r}\psi,K_{rr}),
\end{equation}
\begin{equation}\label{ini_eq_Krr_schem_R}
 \frac{\partial K_{rr}}{\partial r} = \mathcal{K}(\psi,\partial_{r}\psi,K_{rr}) + \epsilon\overline{M}_{K}(\psi,\partial_{r}\psi,K_{rr}),
\end{equation}
which now contains no higher order derivatives on the right-hand side. Higher order derivatives are replaced
by an expansion in $\epsilon$ of lower order derivatives and the equations are now in principle solvable. 
Here, for concreteness we have restricted to first order in $\epsilon$. For the numerical implementation we 
employ finite difference approximations and we numerically integrate through
a Runge-Kutta 4th order. To obtain solutions we perform a shooting procedure in which the value of the fields 
on the inner boundary is found by the implementation of a Newton-Rapson method to ensure that the solutions satisfy the outer 
boundary conditions \eqref{psi_rin} and \eqref{dpsi_rin}.

\subsection{Iterative method, full system (FSII) or order-reduced (ORII)}
The procedure for constructing an iterative solution relies on constructing a solution in terms of an expansion
in $\epsilon$ where, corrections are evaluated with respect to previous iterations. One can choose to solve for
the system of equations \eqref{ini_eq_psi} and \eqref{ini_eq_Krr}---which involve higher derivatives. We refer 
to this as the full system and study its iterative  (or pertubative) integration (FSII). Alternatively, one
can adopt the order-reduced form of the equations~\eqref{ini_eq_psi_schem_R},~\eqref{ini_eq_Krr_schem_R} and 
solve it iteratively (ORII).
We describe the order reduced case (and an analogous method is employed for the FSII case).

First find solutions $\psi_{(0)}$ and $K_{rr (0)}$ for the 
GR equivalent \eqref{ini_reduc}. Then with this zeroth order solution all the components 
of $\widetilde{M}_{\mu \nu}$ can be evaluated to an approximation $\widetilde{M}_{\mu \nu (0)}$. Next one can find solutions 
$\psi_{(1)}$ and $K_{rr (1)}$ for, 
\begin{equation}\label{ini_eq_psi_iter1}
\begin{split}
 \frac{\partial^{2} \psi_{(1)}}{\partial r^{2}} & = -\frac{2}{r}\frac{\partial\psi_{(1)}}{\partial r}- \frac{3}{16}\frac{K_{rr (1)}^{2}}{\psi_{(1)}^{3}}  - \pi \psi_{(1)} \left(\frac{\partial \phi}{\partial r}\right)^{2} 
 -\pi\psi_{(1)}^{5}\Sigma^{2} \\ 
 & + \epsilon\frac{\psi_{(0)}^{5}}{\alpha^{2}}\left(- \frac{\beta^{2}}{4}\widetilde{M}_{rr (0)}  + \frac{\beta}{2}\widetilde{M}_{tr (0)} - \frac{1}{4}\widetilde{M}_{tt (0)}\right), 
\end{split} 
\end{equation}
\begin{equation}\label{ini_eq_Krr_iter1}
\begin{split}
 \frac{\partial K_{rr(1)}}{\partial r} &= -2\psi_{(1)}^{-1}K_{rr (1)}\frac{\partial \psi_{(1)}}{\partial r} - \frac{3}{r}K_{rr (1)} 
 + 8\pi \psi_{(1)}^{4}\Sigma\frac{\partial \phi}{\partial r } \\
 & + \epsilon \frac{\psi_{(0)}^{4}}{\alpha}\left( \beta\widetilde{M}_{rr (0)} - \widetilde{M}_{tr (0)}\right).
\end{split} 
\end{equation} 
This way the terms proportional to $\epsilon$ on  \eqref{ini_eq_psi_iter1} and \eqref{ini_eq_Krr_iter1} 
act simply as source terms in the equations. This procedure can of course be iterated to
obtain $\psi_{(j)}$ and $K_{rr(j)}$.

\subsection{Matter source}
We adopt a largely in-falling scalar field pulse towards a black hole,
\begin{equation}
 \phi(t,r) = \frac{\Phi( u \equiv r +t)}{r},
\end{equation}
with, 
\begin{equation}
\Phi(u) = Au^{2}\exp\left(-\frac{(u-r_{c})^{2}}{\sigma^{2}}\right),
\end{equation}
where $A, r_c$ and $\sigma$ are the amplitude, center and width of the pulse respectively.
Thus, the matter source variables take the following initial values,
\begin{equation}
 \phi_{0}=Ar\exp\left(-\frac{(r-r_{c})^{2}}{\sigma^{2}}\right),
\end{equation}
\begin{equation}
 \Sigma_{0}=\frac{\phi_{0}}{\alpha}\left( \beta\left(\frac{1}{r} - \frac{2(r-r_{c})}{\sigma^{2}} \right) 
 - \left( \frac{2}{r} - \frac{2(r-r_{c})}{\sigma^{2}}\right)\right).
\end{equation}
and as many spatial derivatives of $\phi_{0}$ as required.

\subsection{Boundary conditions}

To solve the initial data equations boundary conditions must be specified. 
In principle, given the high (fourth) order in derivatives of 
the original equations \eqref{ini_eq_psi} and \eqref{ini_eq_Krr}, then up to second derivatives or third derivatives should also be prescribed at the boundaries. However, we have modified these equations
via either the iterative or the order-reduced approaches to get rid of these high order derivatives. 
As a result, one is implicitly specifying these derivatives. In particular, in the order-reduced 
options high order derivatives 
are expressed in terms of lower order ones as in equations \eqref{ini_reduc}, \eqref{ini_3_dev} and \eqref{ini_4_dev}. In the full system iterative integration approach these boundary conditions are redefined at each iteration by means of the previous iteration solution's derivatives.

We explicitly prescribe,
\begin{align}
 &\left. \psi\right|_{r_{out}}= 1 + \frac{M}{2 r_{out}}, \label{psi_rin} \\
 & \left.\frac{\partial \psi}{\partial r}\right|_{r_{out}} = -\frac{M}{2 r_{out}^{2}}, \label{dpsi_rin} \\
 &\left. K_{rr}\right|_{r_{in}}= 0
\end{align}
Also, for simplicity we choose the initial values of the gauge variables to be $\alpha(t=0)=1$ and $\beta(t=0)=0$. This choice simplifies
\eqref{ini_eq_psi}, since $\beta=0$ removes, except from the term proportional to $\widetilde{M}_{tt}$, 
all other modifications to GR. Furthermore, the only non-zero 
modification term in equation \eqref{ini_eq_Krr} is the one proportional to $\widetilde{M}_{tr}$, which vanishes for $K_{rr}=0$. Consequently, since $K_{rr}=0$ as inner boundary condition, $K_{rr}(r<R) \simeq 0$ with
$r=R$ the radius at which the scalar field source is not trivially small. 

\section{Evolution equations \& implementation}\label{EV}
Having presented the evolution equations \eqref{eq_rep_first_order_u}, \eqref{eq_rep_first_order_Pi}
and \eqref{scalar_field} (reduced to first order form via \eqref{eq_rep_first_order_sigma} for convenience) 
we implement
them numerically in the following way. We adopt a method of lines to integrate in time through
a Runge-Kutta 4th order which CFL coefficient set as $dt=0.2\,dx$, where $dt$ denotes the time step 
and $dx$ the spatial (uniform) grid spacing. Our uniform grid extends from $r_i = 0.2M$ to
$r_i = 240M$ and our typical resolution for production runs is $dx=0.019M$.
 Spatial derivatives are discretized via Finite Differences operators
satisfying summation by parts (SBP) (see e.g.~\cite{Calabrese:2003yd,Calabrese:2003vx,MATTSSON2004503,MATTSSON2014432}), of 6th order for inner points and 3rd order at the boundaries and we 
excise the black hole. For reference,
the expression for second and third spatial derivatives satisfying SBP are presented in appendix~\ref{app:derivatives}. We implement Kreiss-Oliger dissipation with operators that are 5th order at the boundary and 6th order at 
the interior points~\cite{Diener2007OptimizedHD}.

\section{Results}\label{results}

\subsection{Initial data}

We now obtain solutions with the three methods described for different values and discuss their
salient features.

\subsubsection{Order-reduced \& full system solutions}
To quantify the performance of the different methods ORDI, ORII, FSII we monitor the 
residual of the original equations \eqref{ini_eq_psi}  and \eqref{ini_eq_Krr} (which requires
evaluating up to fourth order derivatives of the metric) or their order
reduced version (containing up to second order derivatives of the metric). 
We focus first on results obtained with ORDI and ORII. For this set of simulations we take the initial 
scalar field to have  amplitude  $A=1\times 10^{-3}$, to be centered at $r_{c}=20M$ and of width $\sigma=1$.

\begin{figure}[tb]
  \includegraphics[width=\linewidth]{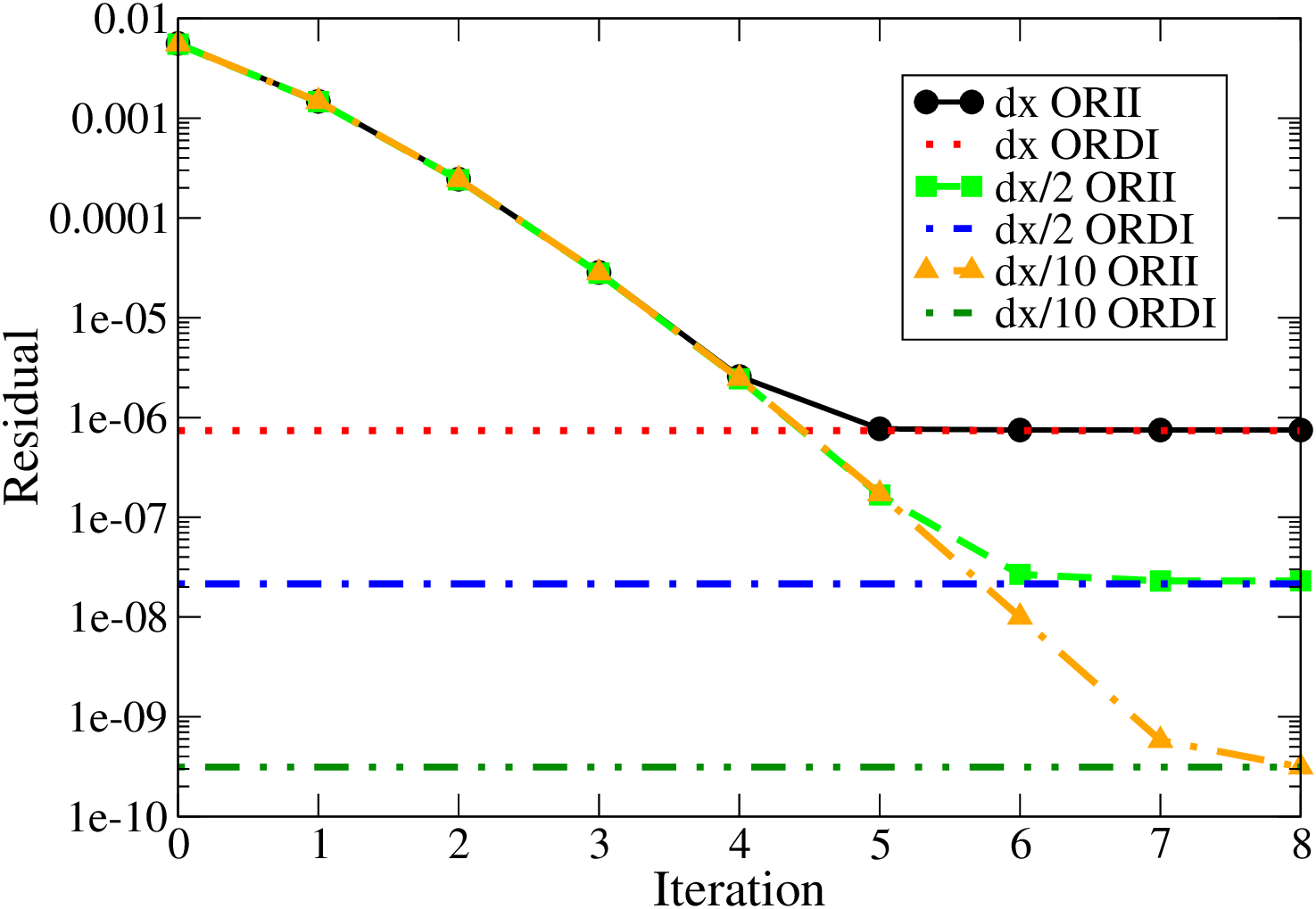}
  \caption{Residuals of equation \eqref{ini_eq_psi} for the iterative solutions as a function of the iteration step. The residuals from the ORDI solutions are represented by the constant horizontal lines for comparison purposes. The different resolutions used are 
 $dx=0.018 M$, $dx/2  $ and $dx/10 $. The residuals of the iterative solutions approach those of the ORDI solutions after sufficient number of iteration steps. (We note in passing the convergence order measured for solutions obtained with the 
ORDI and the ORII methods---for sufficient number of iterations in the latter case---is consistent with the 4th order accuracy of our solver.)}
\label{fig:Residual_Ham_eps_0.01}
\end{figure}

\begin{figure}[tb]
  \includegraphics[width=\linewidth]{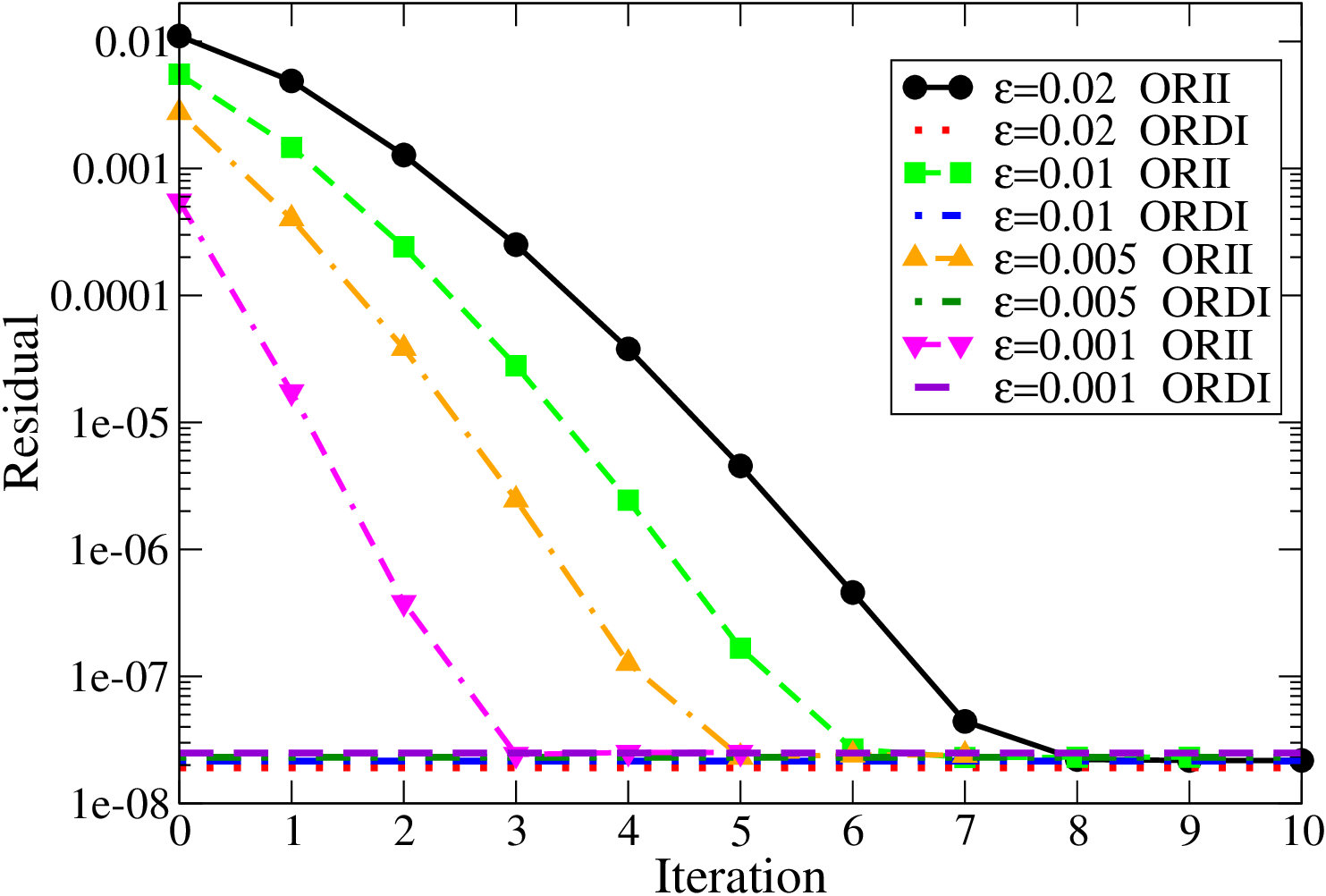}
  \caption{Residuals of equation \eqref{ini_eq_psi} for the iterative solutions with respect to the number of iterations and for a range of 
coupling values for $dx=0.009M$. The residuals from the ORDI solutions are represented by the horizontal dashed lines.}
\label{fig:Residual_Ham_many_eps}
\end{figure}

Figure \ref{fig:Residual_Ham_eps_0.01} displays, for the case $\epsilon=0.01$, the {\em order reduced} 
residual of the
extended Hamiltonian constraint (eqn. \eqref{ini_eq_psi}) for solutions obtained with 
the ORDI and ORII approaches as a function of the
number of iterations performed. The figure shows the results with spatial resolutions
$dx=0.018M$, $dx/2$ and $dx/10$. For the iterated option (ORII), a number of iterations
is required to converge to the solution which, in turns, depends on the spatial resolution. For better resolutions, a larger 
number of iterations is required to achieve such solution.
The ORDI method provides a solution which from the get go gives a residual consistent
with that obtained via the iterative method in the ``asymptotic'' (large number of iterations) regime.
Figure \ref{fig:Residual_Ham_many_eps} shows the same residual but now for different 
values of $\epsilon$ and a single discretization resolution. 
As can be appreciated, a higher number of iterations is required in the
ORII method to obtain the solution for larger values of the coupling. The ORDI method on the
other hand, achieves such solution at once. \\

\begin{figure}[h]
  \includegraphics[width=\linewidth]{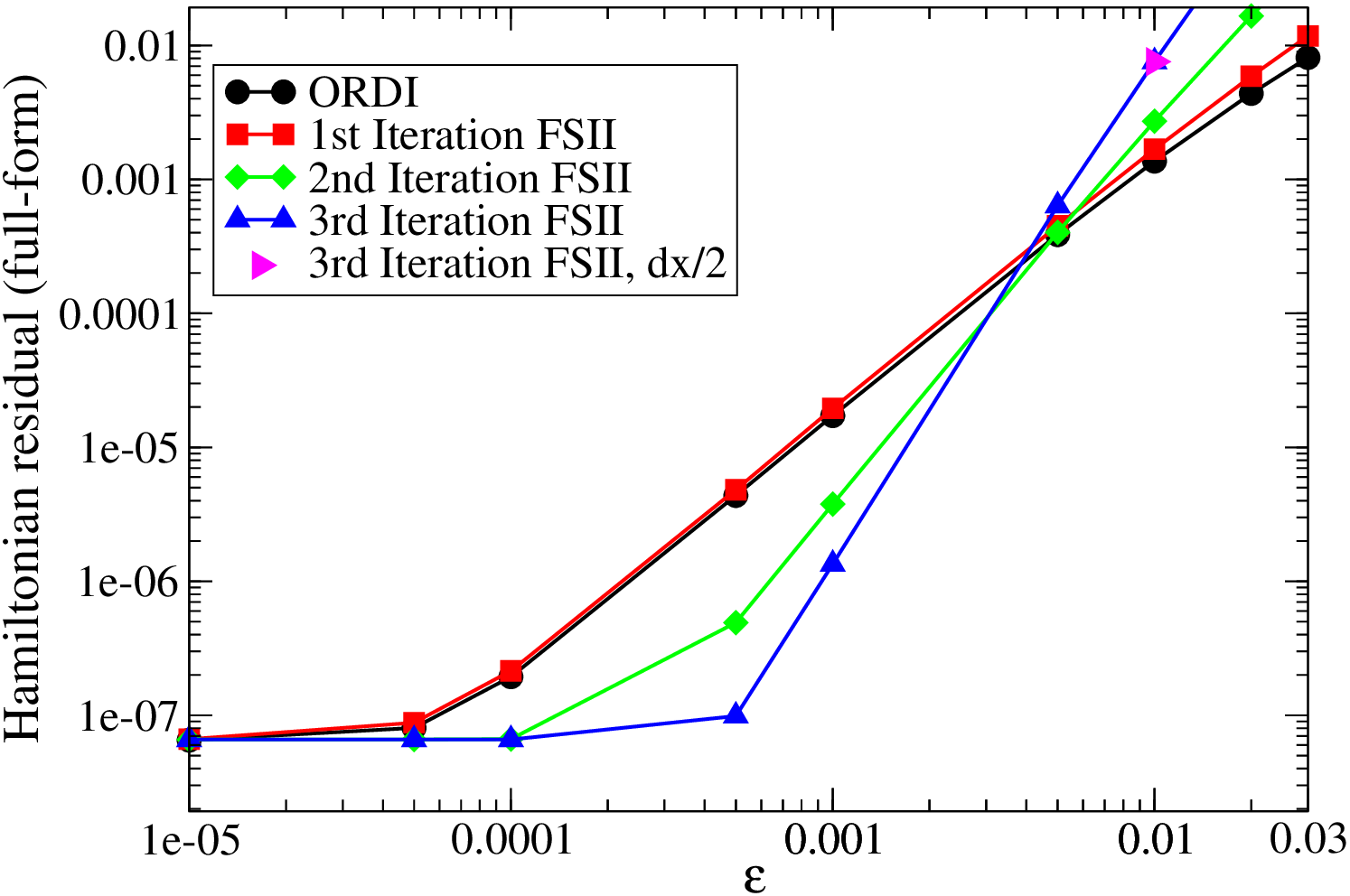}
  \caption{Full-form Hamiltonian residual norm for ORDI and FSII solutions for $dx=0.009M$ as a function of $\epsilon$.} 
\label{fig:Residual_Ham_full_form}
\end{figure}

It is important however to also examine the behavior of the FSII. To that end, we contrast
the norm of the extended Hamiltonian residual---in full form, i.e., not the order reduced one---
equation \eqref{ini_eq_psi} evaluated with the solution obtain with the ORDI and FSII methods.
Figure \ref{fig:Residual_Ham_full_form} shows the residual norm for $dx= 0.009M$ as a function
of $\epsilon$. For small coupling values the residual obtained with the FSII method converges
with a higher power of $\epsilon$ for more iterations, but this behavior degrades as the coupling is
increased. This is a consequence of ``corrections'' to the Hamiltonian in GR becoming too strong
and a related loss of convergence with iteration.
The solution provided by the ORDI method, gives an error consistent with the expected
$\epsilon^2$ behavior as the original Hamiltonian was reduced to such order (as discussed, this can
be formally improved to higher order in a rather direct fashion).

The behavior of residuals obtained from the extended momentum constraint is simple for
our adopted free data. In this equation, the contributions from the extension are non-zero only 
close to the matter sources. Indeed, having chosen an initial scalar field profile supported far from the
black hole the beyond GR terms are significantly smaller than the matter terms. As a consequence, the different
methods provide solutions of similar accuracy (the latter with just one iteration) in all
cases studied. Of course, this behavior need not be true for other boundary conditions, gauge 
choice or location of matter sources.   

From these studies one can draw that in  broad terms the different approaches can be exploited to
obtain solutions reaching comparable accuracy. In the
ORII iterative method, a sufficient number of iterations must be however performed. This number
is dependent on truncation error (i.e governed by $dx$) and physical (i.e $\epsilon$) parameters.
The ORDI method, on the other hand, produces a residual only dependent on truncation error (with respect
to the order reduced form of the constraint).
Finally, the FSII approach yields an increasingly accurate solution in terms of $\epsilon$ for sufficiently
small couplings, but convergence is lost for stronger ones.
We note in passing that these results also provide some sense of the error magnitude that can accumulate 
using a perturbative method during the evolution. Depending on the number of iterations
(or the perturbative order kept),  an error of the order seen in figure \ref{fig:Residual_Ham_many_eps} would
arguably be introduced and its accumulation over the time-length of the simulation can be significant
unless the coupling considered is sufficiently small.\\

Henceforth, we will adopt the solutions obtained with the ORDI method to study the behavior
of perturbed black holes in this theory.

\subsubsection{Solutions' dependence on the coupling parameter}

\begin{figure}[tb]
  \includegraphics[width=\linewidth]{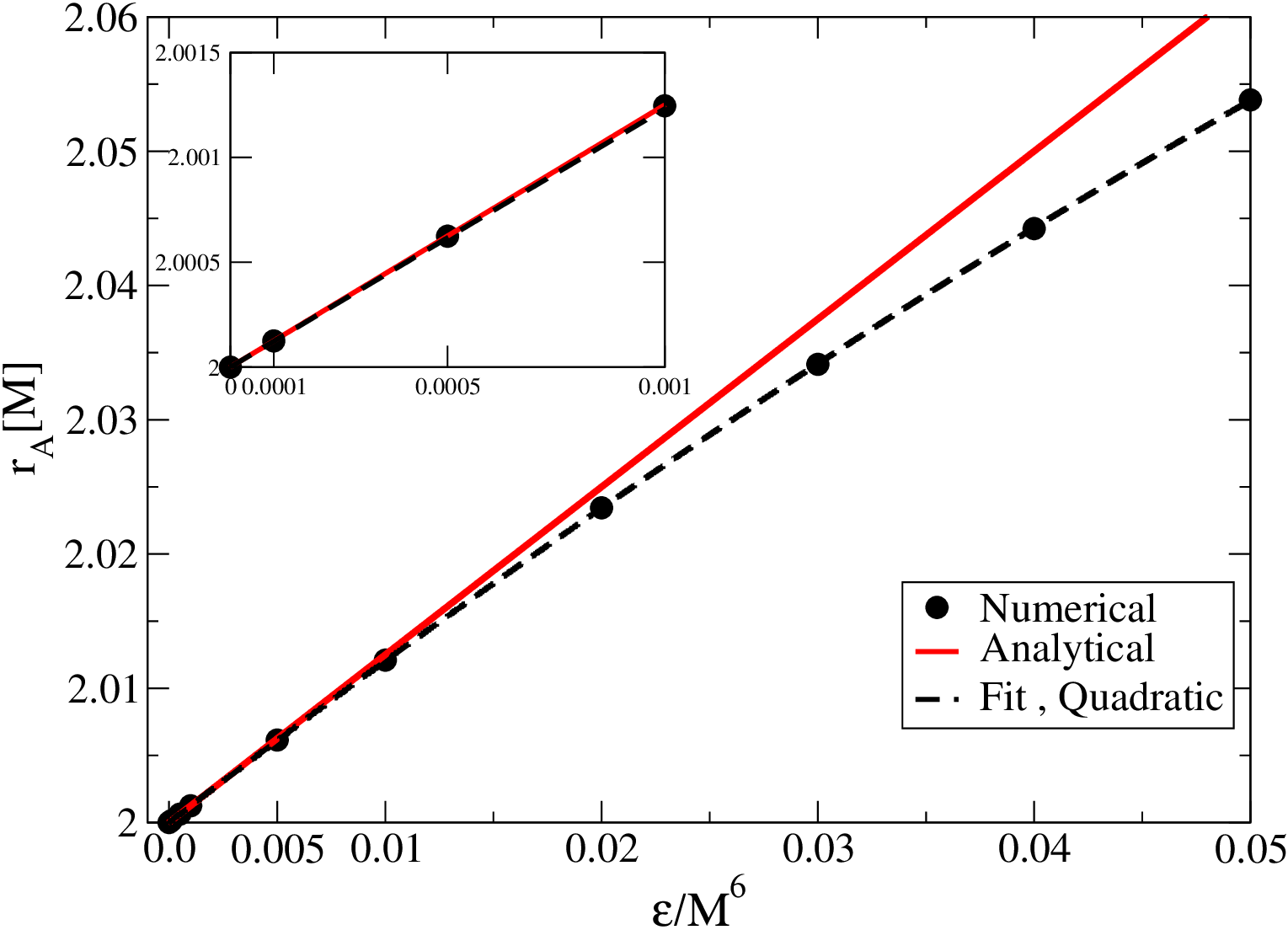}
  \caption{Areal radius of the apparent horizon for different values of the coupling $\epsilon$, for initial data with $A=0$. Black dots correspond
to our numerical solutions, the black dashed line represents the quadratic fit to such data and the red solid line represents 
  the areal radius of the horizon for the analytical solutions found in \cite{PhysRevLett.121.251105}.\vspace{5mm}}
\label{fig:Initial_Area}
\end{figure}

It is informative to examine the dependence of the apparent horizon on the coupling 
parameter $\epsilon$. Figure \ref{fig:Initial_Area} displays 
the apparent horizon areal radius and its change as the coupling $\epsilon$ is increased for
our initial data. The figure 
shows the value of the areal radius of the apparent horizons found numerically with our solutions as well as with 
the analytical (perturbative) solutions found in \cite{PhysRevLett.121.251105} as a function of $\epsilon \, M^{-6}$. A fit to our 
numerical data of the form $r_A^{H}= 2M + s\, \epsilon \, M^{-6} + q\, \left(\epsilon \, M^{-6}\right)^{2}$
gives $s=1.234$ and $q=-3.179$. This is in agreement with the expression obtained from
the analytical (linear) solution where $s_{analytical}=1.25$. Recall that while the equations
giving rise to the solution of  \cite{PhysRevLett.121.251105} and ours are linear in $\epsilon$, the solutions will differ at higher
orders due to boundary conditions and our solution with the ORDI method which, in essence provides a 
resummed solution. Thus, differences at
order $\epsilon^2$ are expected.
Figure~\ref{fig:Initial_Area} illustrates both curves; as $\epsilon$ increases, the quadratic contribution 
leads the numerical solutions peeling off the analytical ones, respect to the apparent horizon radius, though the difference is smaller than $3\%$ for 
$\epsilon = 0.05 M^6$. 

\begin{figure}[tb]
  \includegraphics[width=\linewidth]{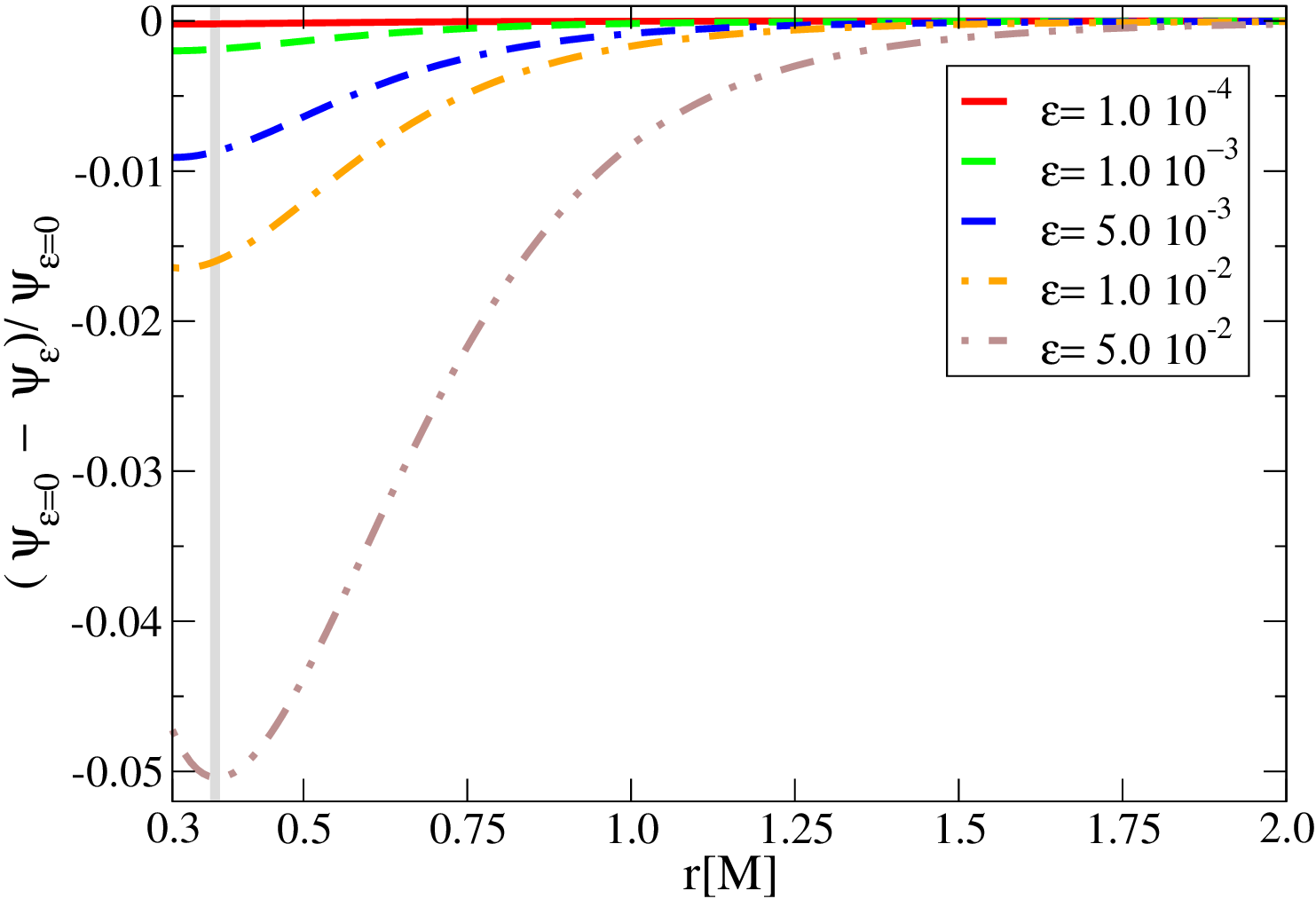}
  \caption{Relative difference $({ \psi_{\epsilon=0} - \psi_{\epsilon} })/{\psi_{\epsilon=0}}$ for different values of the coupling parameter $\epsilon$ as a function of the coordinate radius $r$. The gray vertical line
is included to guide the eye, giving an indication of the apparent horizon locations( which vary 
with $\epsilon$) .}
\label{fig:Relative_psi}
\end{figure}

To get a sense of the differences (magnitude and radial dependence) introduced by the correcting terms,
figure \ref{fig:Relative_psi} shows the relative difference between the conformal factor $\psi$ obtained for different values of $\epsilon$ and the GR solution ($\epsilon=0$). Departures from the GR solution, while very small asymptotically, become larger as the radius decreases reaching values 
above $1\%$.

\subsection{Dynamical behavior}
We now turn our attention to the dynamical evolution of a (mainly) incoming self-gravitating scalar
field configuration with different choices for its amplitude $A$ together with 
several values for the coupling parameter $\epsilon$. The values of $dx=0.019M$, $\sigma=0.018M$ and $r_{c}=20M$ are fixed for all the results presented in this section.   \\

As the evolution proceeds, a common qualitative behavior is seen in all cases; namely, much of the scalar field falls towards the black hole interacting with it
while a small portion of the initial scalar field leaves the computational domain in a short
time (afterwards, the resulting spacetime
has an asymptotic mass $M_{as}=0.9998$ in the domain explored by the numerical implementation).
To provide a quantitative understanding of the ensuing dynamics,
we focus on the behavior of the: {\em apparent horizon}, {\em quasi-normal behavior of the scalar radiation} and {\em suitable geometric invariants}.

\subsubsection{Apparent horizon}
As the scalar field falls into the black hole, the area of the event horizon (an thus its mass) grows but
a closer inspection reveals a subtle and a-priori unexpected dependence on $\epsilon$.

\begin{figure}[tb]
 \includegraphics[width=\linewidth]{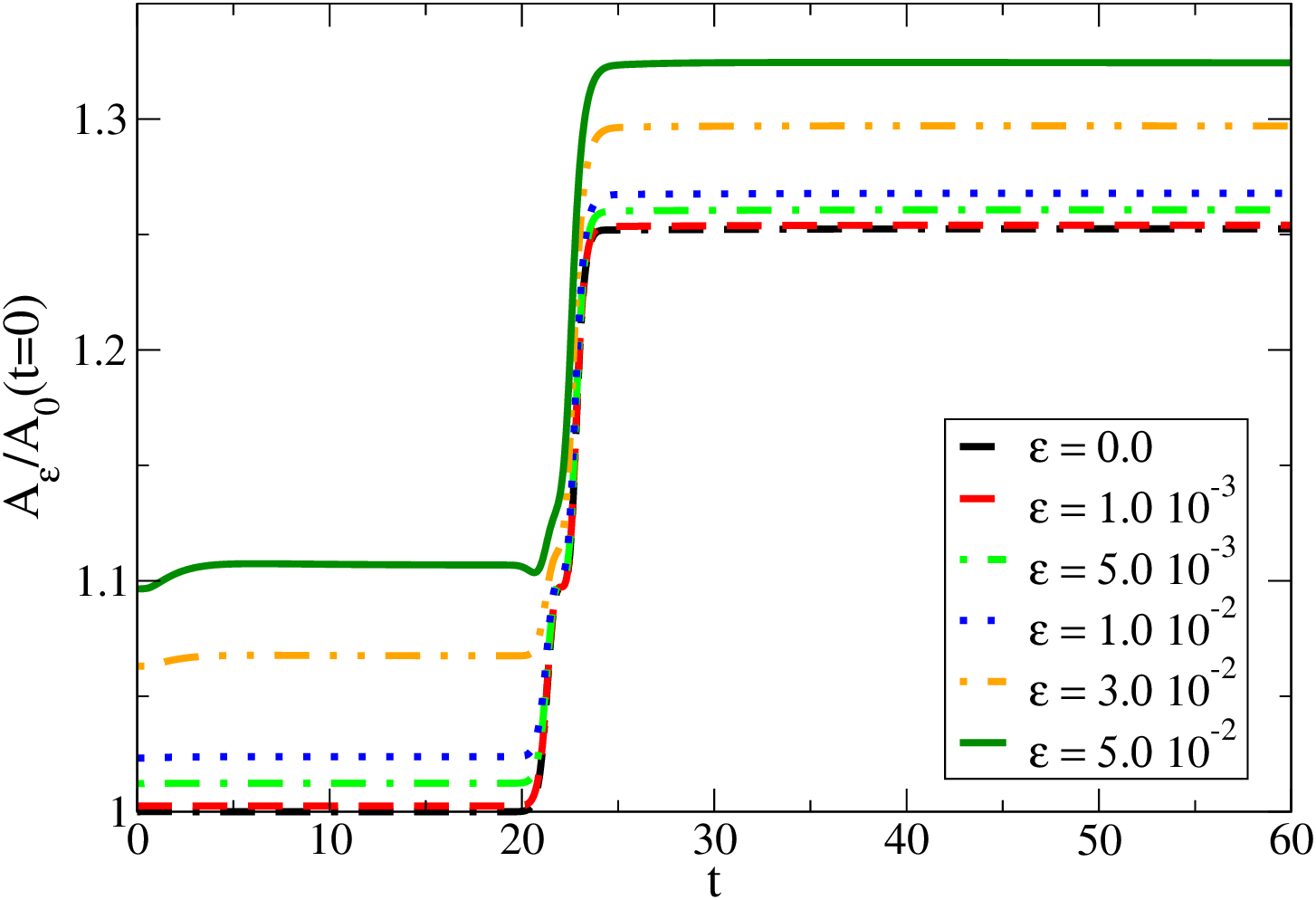}
  \caption{Area of the apparent horizon as a function of time for different values of $\epsilon$. All curves are normalized by 
the corresponding initial area in GR (i.e $\epsilon=0$). The initial irreducible
  mass for all cases is $M_{i}=0.8933$, the final $M_f{}=0.9998$ and the total mass of the spacetime $M=1.0$. }
\label{fig:BH_areas_time}
\end{figure}

Figure \ref{fig:BH_areas_time} 
shows the apparent horizon area (normalized by the initial area in the $\epsilon=0$ case) as a function of time for different values of the coupling parameter $\epsilon$. All of the 
simulations used to make Figure \ref{fig:BH_areas_time} present a black hole with initial irreducible mass $M_{i} =  0.8933$ a 
final mass of $M_{f}=0.9998$, while the initial mass of the full spacetime is $M=1.0$ and the amplitude of the scalar pulse is $A=1 \times 10^{-3}$.

The overall behavior for all of these curves is similar; namely, the horizon grows as scalar field
energy is accreted until it reaches an approximately stationary state describing a black hole 
with a mass up to
$\approx 12\%$ larger.  The case with $\epsilon= 0$, as expected, gives rise to
a non-decreasing behavior of the apparent horizon area. However, subtle details can be seen 
with  $\epsilon \neq 0$ which are more marked for larger values of the
coupling parameter.

\begin{figure}[tb]
 \includegraphics[width=\linewidth]{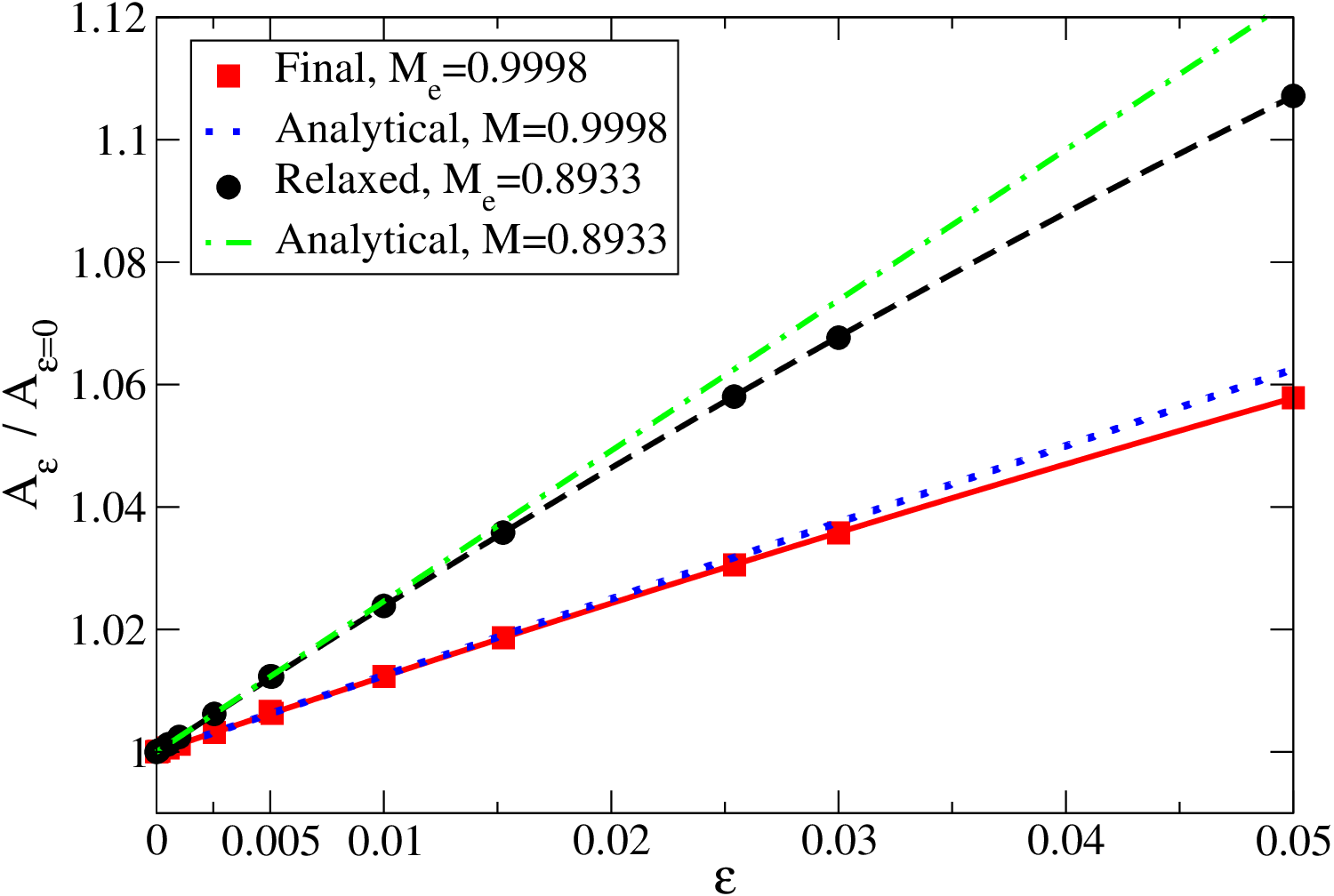}
  \caption{$A_{\epsilon}/A_{0}$ as a function of $\epsilon$ at two particular instances. Red squares denote the late time solution and solid red line  its quadratic fit. 
  Black circles correspond to the intermediate solution and the dashed black line its quadratic fit. The dotted blue and dashed and dotted green lines correspond to the analytical 
  solutions of \cite{PhysRevLett.121.251105} for $M=0.9998$ and $M=0.8933$ respectively.}
\label{fig:Horizons_relaxed}
\end{figure}

First, one observes an initial transient growth in the apparent horizon area even though no scalar field energy has been accreted. This behavior is not surprising however, as it related to the initial data adopted which is 
non-stationary. 
The future development of the initial data, after the transient stage
reveals a transition to a new intermediate (i) stage when the apparent horizon area does not 
change until the (main) accretion stage ensues. 
At late times, the solution is described by an essentially stationary final (f) configuration. 
The asymptotic state described by the apparent horizon (and thus an excellent approximation to the event horizon), can be  understood by 
computing the fraction $A_{\epsilon}(M_{f})/A_{0}(M_{f})$ of the black hole and compared it to the fraction $A_{\epsilon}(M_{i})/A_{0}(M_{i})$ 
at the initial time, or  with the area during the intermediate stage. In figure \ref{fig:Horizons_relaxed} we show 
these quantities as well as the one corresponding to the analytical solution from \cite{PhysRevLett.121.251105} as a function of $\epsilon$.
As this figure shows, the curves for intermediate and late time solution match the curve for the 
analytical solution at small couplings and for the corresponding masses.
Indeed, a quadratic fit of the form $A_{\epsilon}/A_{0}= 1 + s\, \epsilon \, M_e^{-6} + q\, \left(\epsilon \, M_e^{-6}\right)^{2}$ to our 
data to the area gives,
$s_{i}=1.251$ and $s_{f}=1.252$ (here $M_e$ is the irreducible
mass estimated during the intermediate and final stages respectively: $M_e=0.8933$ or $M_e=0.9998$). Both these values agree with that
of the analytical (linear) solution $s_{a}=1.25$. Furthermore the obtained values of $q_{i}$ and $q_{f}$ are also consistent with each other.

Second, and at first-sight surprising, one sees a momentary {\em small decrease} in the area of the apparent horizon as the scalar field interacts with it; this behavior is more marked for larger
values of $\epsilon$. This effect, when seen through the lense of GR can be traced to the failure
of the null convergence condition (NCC). In such cases, the area of the event horizon---and hence 
that of the apparent horizon---can decrease in size~\cite{Hawking:1973uf,Hayward:1993mw,Ashtekar_2004}.

\begin{figure}[tb]
 \includegraphics[width=\linewidth]{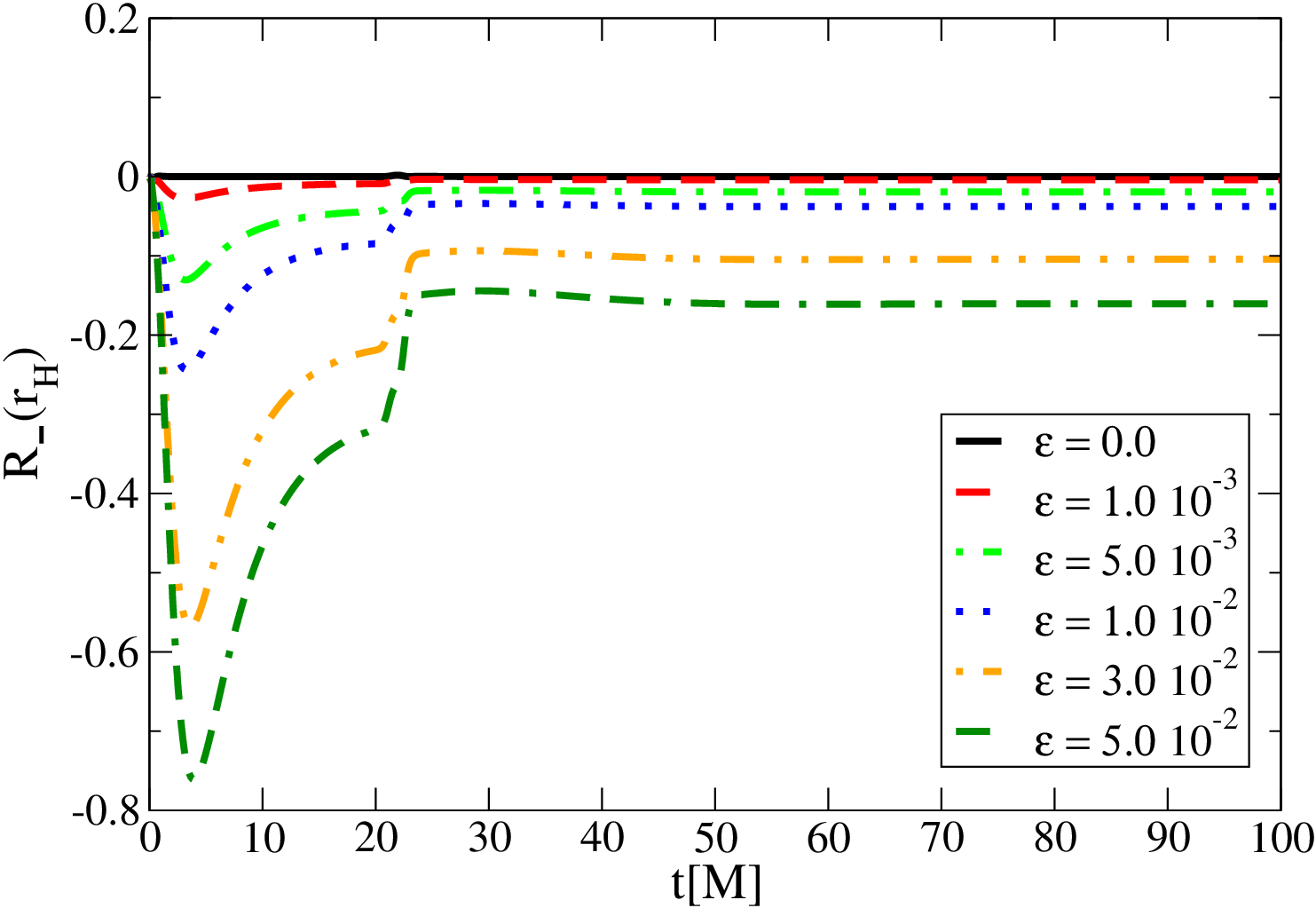}
  \caption{$R_{-}$ evaluated at the apparent horizon as a function of coordinate time $t$ for several values of $\epsilon$.\vspace{7mm}}
\label{fig:NC_minus_at_horizon}
\end{figure}

To examine the NCC we monitor $R_{\pm}\equiv R_{\alpha \beta} k_{\pm}^{\alpha}k_{\pm}^{\beta}$, 
  where $k_{\pm}^{\alpha}$ are the only (up to multiplicative factors) future directed null vectors present in spherical symmetry. Their expressions are given by: 
 \begin{equation}
  k^{\alpha}_{\pm} = \left (1 ,-\beta \pm \frac{\alpha}{\sqrt{g_{rr}}} ,0,0 \right)
 \end{equation}
Figure \ref{fig:NC_minus_at_horizon} shows the value of $R_{-}$ evaluated at the apparent horizon as a function of time for several values of $\epsilon$. Clearly, 
 the NCC is being violated at all times for the $\epsilon \ne 0$ solutions, and this violation becomes more marked as the coupling $\epsilon$ increases.  

\begin{figure}[tb]
 \includegraphics[width=\linewidth]{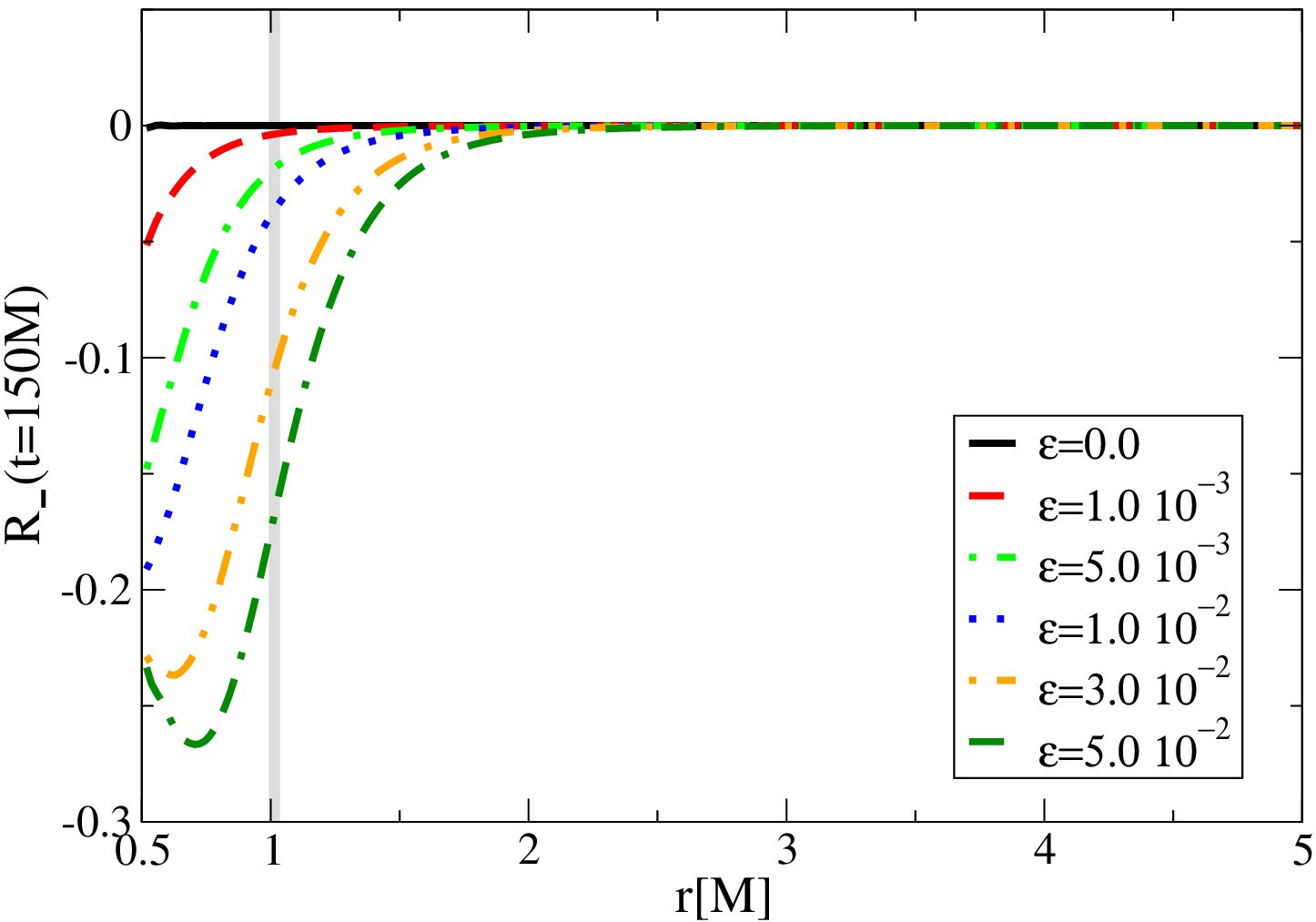}
  \caption{$R_{-}$ at $t=150M$ as a function of coordinate radius $r$ for several values of $\epsilon$. The gray vertical line
is included to guide the eye giving an indication of the apparent horizon locations.)}
\label{fig:NC_minus_at_t150M}
\end{figure}

Figure \ref{fig:NC_minus_at_t150M} presents a snapshot of $R_{-}$ as a function of coordinate radius $r$ at coordinate time $t=150M$ for different values of $\epsilon$. The NCC violations 
are not only present in the vicinity of the apparent horizon, but they persist in the whole spatial domain.

Similar results are found for $R_{+}$, for which the NCC is as well violated. The violation of the
NCC also stresses that the dynamics within extensions to GR can 
display surprising phenomena that must be understood for potential implications on gravitational wave data.

\subsubsection{Ringing and QNM}

\begin{figure}[tb]
 \includegraphics[width=\linewidth]{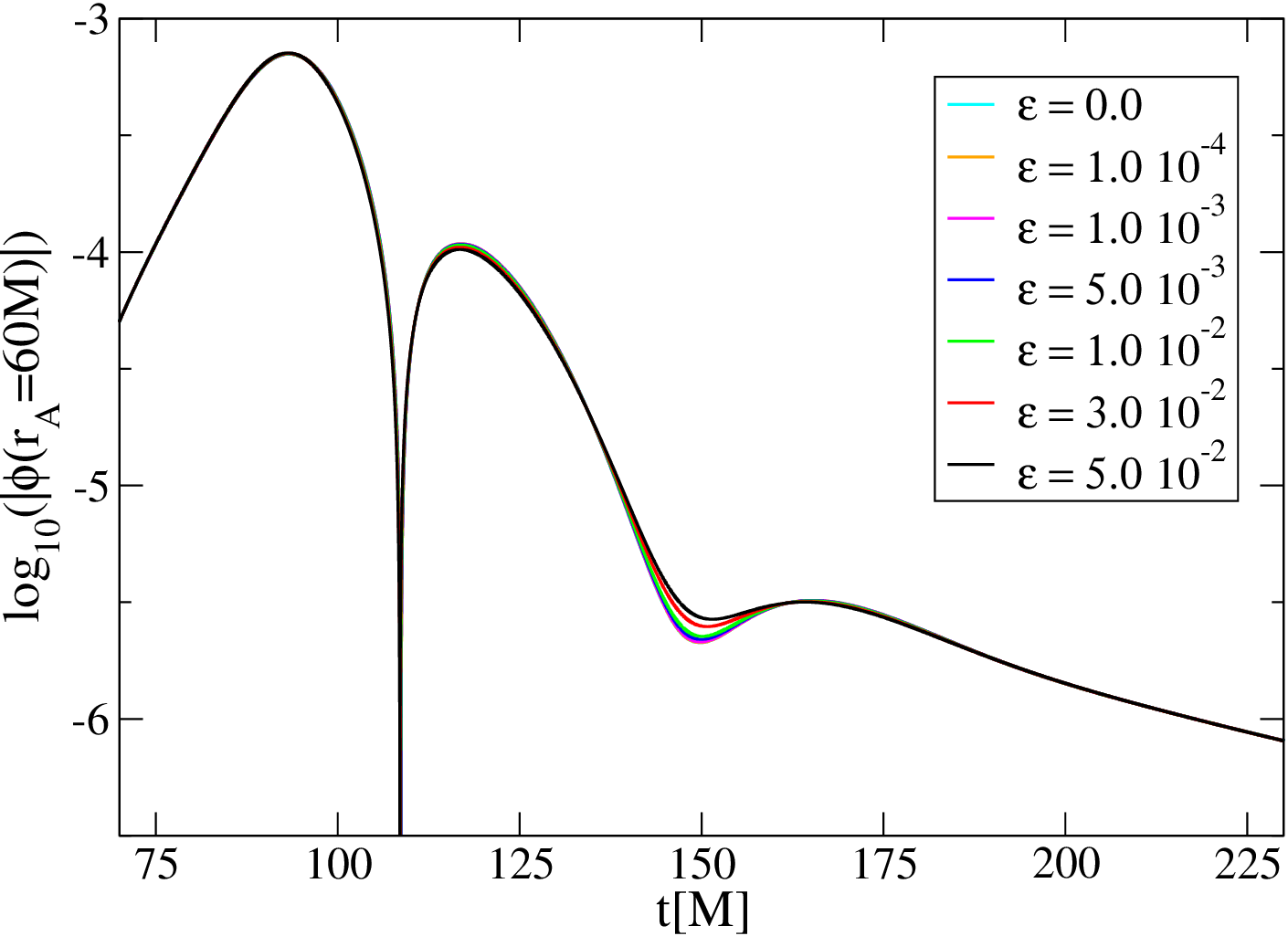}
  \caption{$\log(|\phi|)$ at areal radius $r_{A}=60M$ as a function of coordinate (i.e. asymptotic) time for a wide range of values of $\epsilon$. 
}
\label{fig:logphi}
\end{figure}

In GR, the (linearized) study of perturbed black holes reveals a quasi-normal behavior where the radiation fields (scalar, vector or tensor modes)
are largely described by a set of exponentially decaying oscillations with decay rate and oscillation frequency tightly tied to the black hole parameters
(mass and angular momentum). While the existence of analog modes for black holes in generic 
EFT-motivated theories has not been rigorously analyzed, at an
intuitive level a similar behavior is expected if black holes in such theories (and within the EFT regime) are stable\footnote{After all,
perturbations are described still by propagating waves in a leaky cavity---loosing energy
into the black hole or radiated to infinity---and the spacetime is described by a small set of parameters $\{M,J,\epsilon\}$ which would determine
the decaying/oscillatory behavior.}. We here study this behavior for the scalar field in spherical symmetry ($l=m=0$) which we fit to a behavior
given by,
\begin{equation}
\phi(t,r) = \sum_{n=0}^{\infty}c_{n}\exp({i\omega_{n}(t-r)}),
\end{equation}
where $\omega_{n}$ are complex frequencies and $n$ is the overtone index. As expected,
a behavior akin to the familiar quasi-normal ringing is observed as can be appreciated in figure \ref{fig:logphi}
which shows the  scalar field behavior at a large distance vs time. The field is dominated by presence of damped oscillations 
 with a subtle dependence on $\epsilon$. This figure also shows a transition between a QNM behavior and a power-law tail dominated one. The power law exponent that we observe on this curves is $t^{-3}$ and thus 
consistent with analytical and numerical predictions for this mode in the GR case.

\begin{figure}[tb]
 \includegraphics[width=\linewidth]{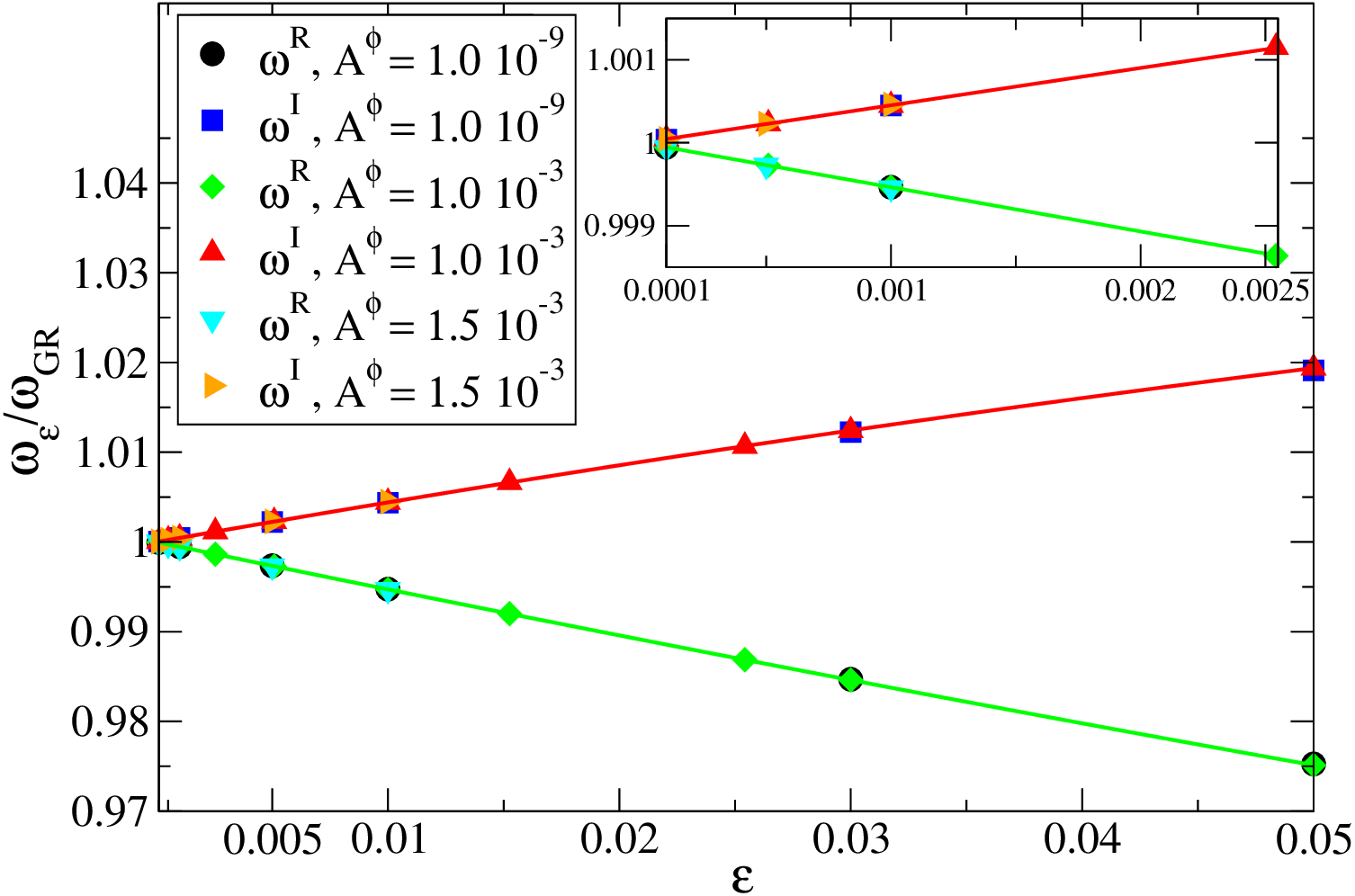}
  \caption{Relative QNM frequency for the strong and weak field
cases as a function of $\epsilon$. The solid lines in the plot are quadratic fits to the numerical data. The obtained parameters 
  are in agreement between the weak and strong field cases.}
\label{fig:QNM}
\end{figure}

For a quantitative analysis, we focus on the least damped $l=m=n=0$ mode. We extract
the value of the field at an areal radius $r_{A}=60M$ for three cases defined by initial
amplitudes of the scalar field,  a weak one of $A=10^{-9}$, and two strong ones
with $A=10^{-3}$ or $A=1.5 \, 10^{-3}$) centered initially at coordinate radius
$r= 20M$ and width $\sigma=1.0M$ . In the strong field cases the final mass of the black hole increases by $\approx 12\%$ and $\approx 32 \%$ respectively after accretion. We extract both the real, $\omega^R$, and imaginary $\omega^I$  
frequencies and focus on their dependence on the coupling parameter $\epsilon$.
Figure~\ref{fig:QNM} illustrates our results taking the ratio of the obtained values with respect
to the ones for the GR case. The QNM frequencies values obtained for the GR simulation ($\epsilon=0$) are $\omega^{R}= 0.109$ and $\omega^{I}=0.104$  and 
are within $1\%$ from the known values predicted by linear perturbation theory \cite{PhysRevD.34.384,2009CQGra..26p3001B}. As can be appreciated in the figure, there is a somewhat
larger deviation for $\omega^R$ than for $\omega^I$. A general, simple, quadratic fit for both cases is,
\begin{eqnarray}
\omega^R &=& \omega^R_{GR} (1 - 0.54\epsilon + 0.77 \epsilon^{2} ), \\
\omega^I &=& \omega^I_{GR} (1 + 0.45\epsilon -1.33 \epsilon^{2} ).
\end{eqnarray}
Furthermore we have observed that this scaling is independent of the initial amplitudes $A$ of the scalar field and of the timescale $\tau$ introduced in equation \eqref{eq_rep_first_order_Pi}. 
We have found that this scaling is in good agreement with the analytical study of QNM frequencies for black holes in higher derivative 
theories ~\cite{2020arXiv200503671C} (including the one studied here). In our notation their predictions translate to:
\begin{eqnarray}
\omega^R_{analytical} &=& \omega^R_{GR} (1 - 0.503\epsilon  ), \\
\omega^I_{analytical} &=& \omega^I_{GR} (1 + 0.484\epsilon  ).
\end{eqnarray}
The discrepancy in the correcting factor is $\approx 7\%$ between our numerical prediction and their perturbative, analytical treatment. 

\subsubsection{Curvature invariant}

As a final step, we monitor the scalar curvature invariant $\mathcal{C}\equiv R^{\alpha \beta \gamma \delta}R_{\alpha \beta \gamma \delta}$ to obtain further insights on the spacetime. 

\begin{figure}[tb]
 \includegraphics[width=\linewidth]{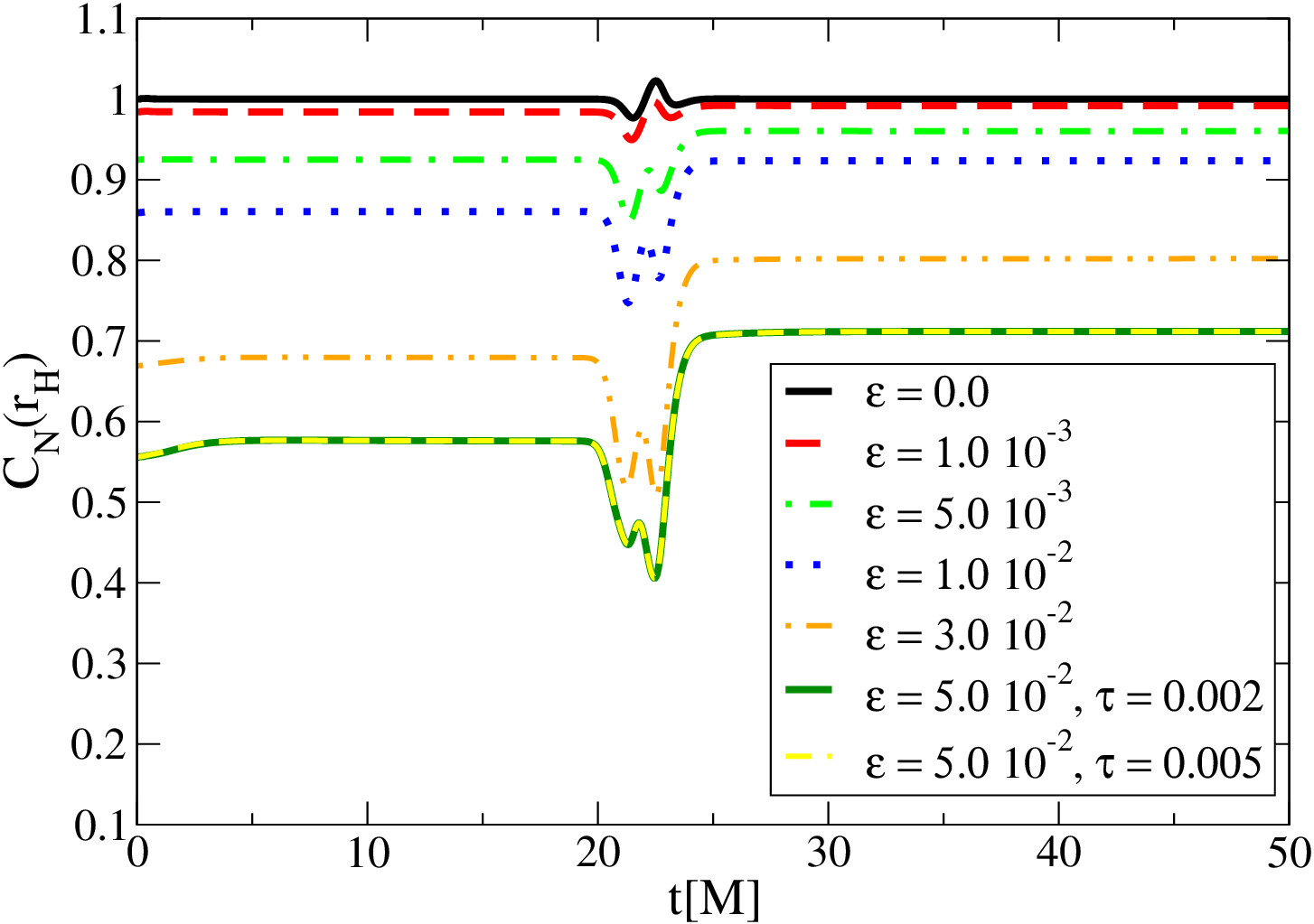}
  \caption{$\mathcal{C}_{N}\equiv 4/3 \, \mathcal{C}M_{H}^{4}$ evaluated at the apparent 
  horizon as a function of $t$ for different values of $\epsilon$. The case $\epsilon=0.05$ is also presented with a longer time scale $\tau=0.005$.}
\label{fig:Kret_at_H}
\end{figure}

Figure~\ref{fig:Kret_at_H} shows the value of $\mathcal{C}_{N}\equiv 4/3\, \mathcal{C}M_{H}^{4}$ 
(normalized this way  as $\mathcal{C}_{N}=1$ for a Schwarzschild black hole) evaluated at the apparent horizon as a function of time for different values of the coupling parameter $\epsilon$ (where $M_{H}$ is the 
irreducible mass of the apparent horizon, an $\epsilon$-dependent quantity in this theory). Note that the $\epsilon = 0.0$ curve departs from $1$ only around the time 
 when the black hole is accreting the scalar pulse and the local solution is not described by the Schwarzschild geometry. For non-zero coupling values,  $\mathcal{C}_{N}$ departs further away from 1 as $\epsilon$ increases.
Since the black hole grows via accretion, the difference with respect to the value for Schwarzschild decreases after it grows
as corrections in the theory are governed by curvature.  Turning our attention to the transient (accreting) stage, fluctuations induced by accretion vary strongly with $\epsilon$, both in amplitude and functional dependence. This indicates interactions of the black hole 
and the scalar field are strongly modified in this theory. 
The figure also includes two curves ($\tau=0.002$ and $\tau=0.005$) for the strongest coupling case ($\epsilon=0.05$) 
to illustrate our results are independent of the timescale $\tau$.

\begin{figure}[tb]
 \includegraphics[width=\linewidth]{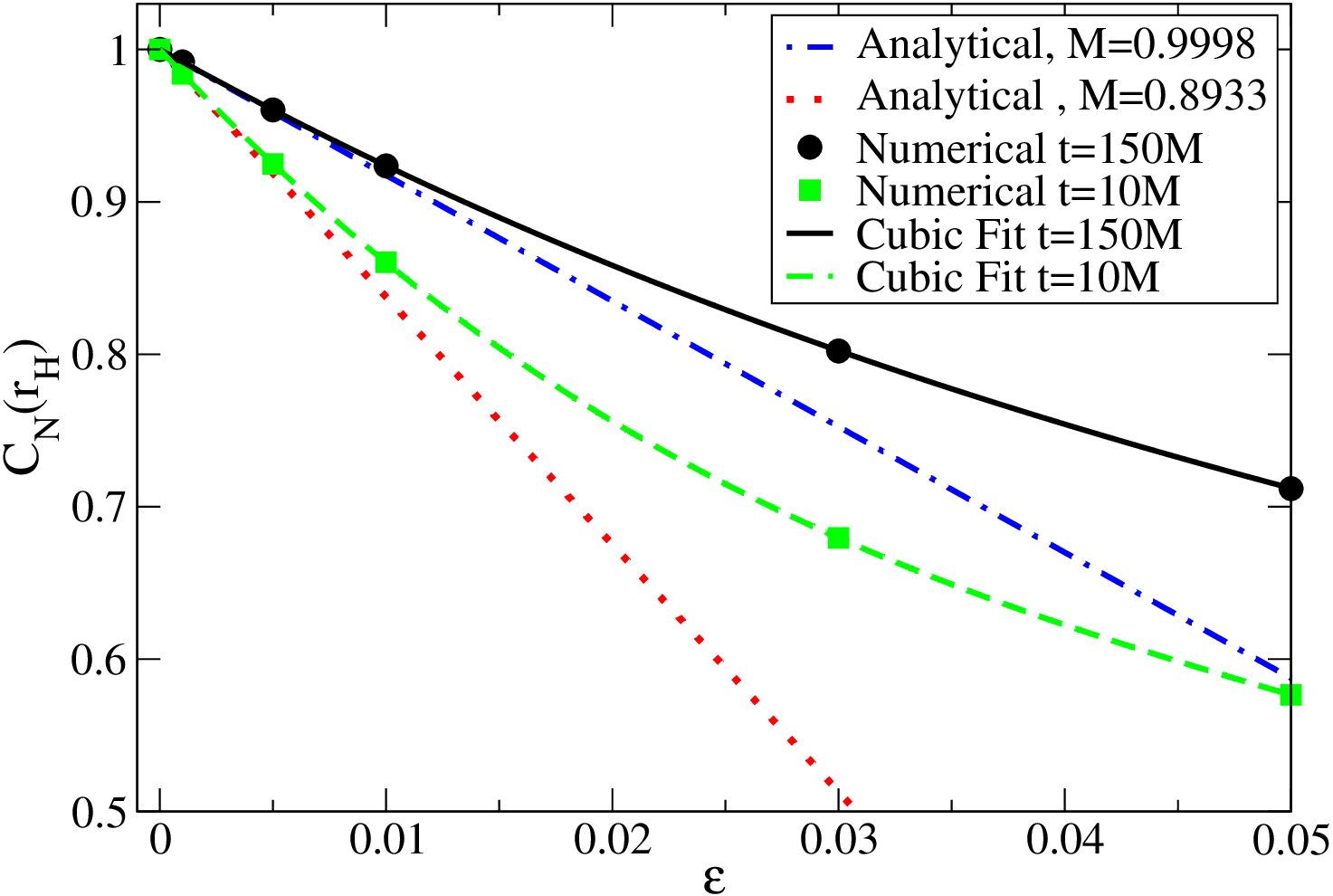}
  \caption{Green squares and black dots represent the value of $\mathcal{C}_{N}$ at the horizon for two coordinate times $t=10M$ and $t=150M$ as a function of $\epsilon$. The green dashed line and 
  the black solid lines are cubic fits to our numerical data. The dashed and doted blue and doted red lines display the prediction of $\mathcal{C}_{N}(r_{H})$ from the solutions obtained in~\cite{PhysRevLett.121.251105} for masses of $M=0.9998$ and $M=0.8933$ respectively.}
\label{fig:Kret_hor_anal_num}
\end{figure}

Figure~\ref{fig:Kret_hor_anal_num} shows the values of $\mathcal{C}_{N}$ (evaluated at the apparent horizon) as a function of $\epsilon$ at two particular 
times, $t=10M$  and $t=150M$, that describe black holes that are approximately stationary during
the intermediate and final stages. Additionally, we include the analytical value computed with 
the black hole solution found on ~\cite{PhysRevLett.121.251105} and fits to our numerical values. The most evident feature of this figure is the clear departure of our numerical solutions from the linear 
result  
(which gives by $\mathcal{C}_{N} = 1 - {33}/{4}\,\epsilon (M^{GR}_{H})^{-6}$ (where $M^{GR}_{H}$ is the irreducible mass of the black hole in the $\epsilon=0$ case.) for large 
enough values of $\epsilon$. Performing a cubic fit of the form 
$\mathcal{C}_{N}=1 +s\, \epsilon (M_{e})^{-6} + q\, (\epsilon (M_{e})^{-6})^{2} + c\, (\epsilon (M_{e})^{-6})^{3}$ 
to our data points, the fitted values 
of the linear term coefficient for the $t=10M$ and $t=150M$ solutions are $s=-8.22$ and $s=-8.23$ 
respectively. The results for the linear coefficients are in  in good agreement with the value $s=8.25$ 
obtained with the analytical solution. We note that if $\mathcal{C}_{N}$ is plotted as a function of $\epsilon (M_{e})^{-6}$ then the curves drawn for $t=10M$ and $t=150M$ match to an excellent degree.

\begin{figure}[tb]
 \includegraphics[width=\linewidth]{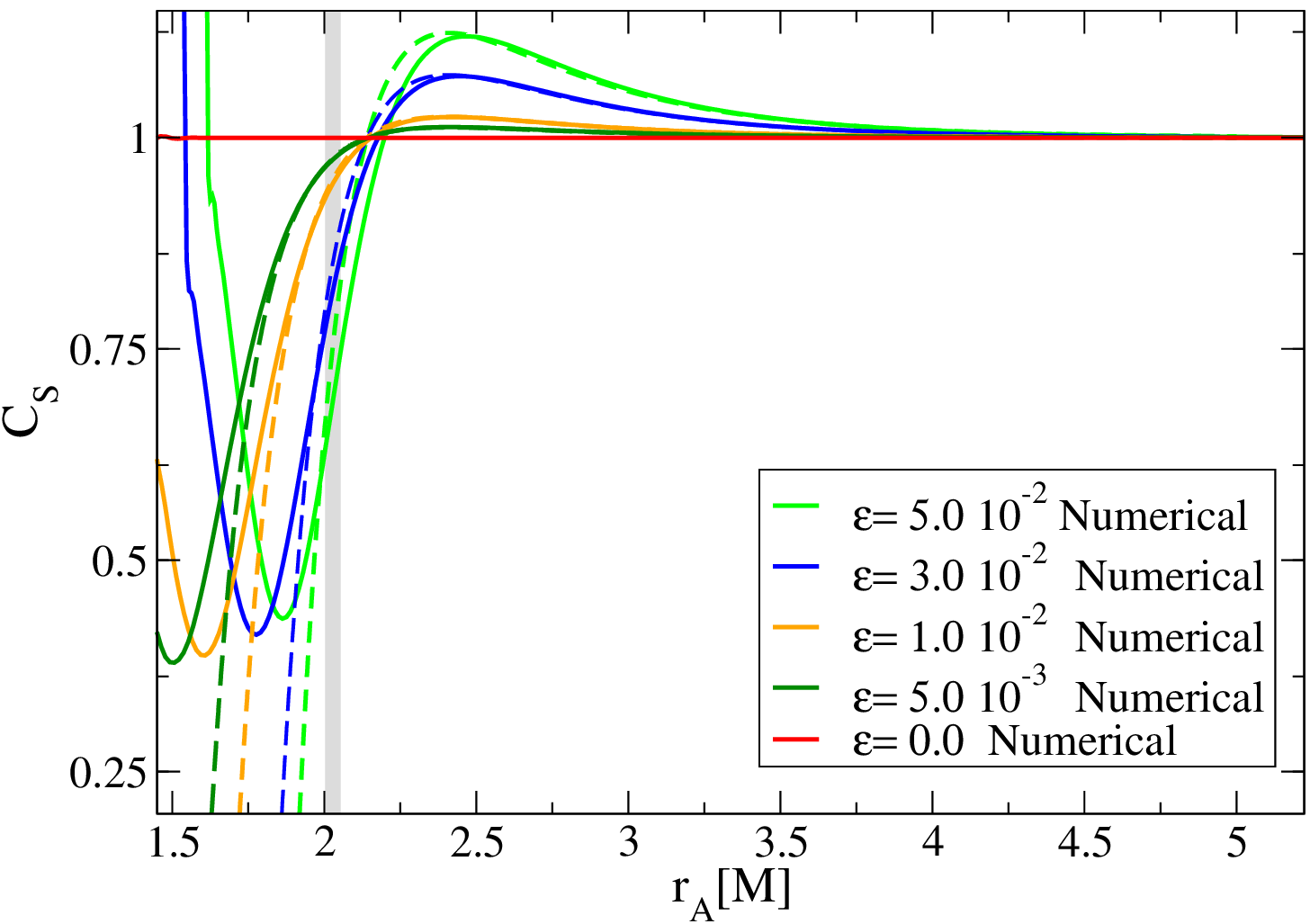}
  \caption{$\mathcal{C}_{S}$ as a function of areal radius $r_{A}$ for different values of $\epsilon$. The solid lines represent our numerical solutions for $t=150M$ (describing a black hole of mass $M_{GR}=0.9998$), while 
  the dashed lines correspond to the analytical predictions of $\mathcal{C}_{S}$ from the solution in~\cite{PhysRevLett.121.251105}. The gray vertical line
is included to to guide the eye giving an indication of the apparent horizon locations.}
\label{fig:Kret_r}
\end{figure}

 In figure \ref{fig:Kret_r} we show the behavior of of $\mathcal{C}_S \equiv\mathcal{C}{r_{A}^{6}}/{(48\, M_e^2)}$ as function of the areal radius $r_{A}$ for $t=150M$  
 for a a wide range of $\epsilon$ values along with the linear analytical predictions for this quantity. At far distances from the
black hole, all curves approach the value $1$ expected for a Schwarzschild black hole as expected---since corrections decay at
a high rate with distance. Close to the black hole however, the quantity
peels off from the Schwarzschild value and while such behavior is more marked---inside the black hole---it is non-trivial in
its outer vicinity.

%%%%%%%%%%%%%%%%%%%%%%%%%%%%%%%%%%%%%%%%%%%%%%%%%%%%%%%%%%%

\section{Final Comments}\label{finalcom}

In this work we illustrated the implementation of a method to control
the presence of higher derivative terms in extensions to GR.
Using reduction of order techniques, we traded higher-time derivatives to
eliminate Ostrogradsky's type ghosts and through the use of the ``fixing equations''
method~\cite{Cayuso:2017iqc} we controlled higher-spatial derivatives. This combined approach
allows us to treat highly complex non-linear theories with higher derivative
contributions in a non-iterative fashion (which can also be referred to as non-perturbative 
from the point of view of how correcting terms are handled. See e.g.~\cite{Okounkova:2019dfo}
for such a perturbative approach).

We illustrated the benefits of proceeding this way by studying the dynamics of
a self-gravitating scalar field in a spherically symmetric black hole spacetime
within a theory displaying derivatives up to 4th order and corrections
to GR (with combined gradient contributions of order $\lambda^{-8} \Lambda^{-6}$). 
We described how initial data can be constructed directly integrating the resulting constraint equations
and contrasted the solution with those obtained with iterated/perturbative approaches.  
Our results demonstrate that, for sufficiently weak couplings the solutions agree, but for larger
ones there is increasing disagreement and a larger number of iterations might be required to achieve a
sufficiently small residual. In particular, this observation gives a sense of the potential size of
 error that would be incurred, and accumulated, in dynamical studies utilizing an iterated approach
restricted to just the first correction (or treated perturbatively to first order).

Studying the future development of data describing a scalar field perturbing a black hole in such
theories, we observed the apparent horizon can reduce in size due to the NCC
being violated. As well, the scalar field displays a QNM behavior reminiscent
of that familiar in GR, but with decaying and frequency rates that differ more strongly for larger couplings.
In particular we find that  relative differences in decay rate and oscillatory frequency 
scale as $\simeq 0.5 \epsilon$ which, if translated in similar fashion
to the gravitational wave sector would imply useful constraints (or detection!) could be placed by
upcoming detections. These results can help further inform approaches to parameterize deviations from GR signals  by
making explicit connections with putative theories
e.g.~\cite{Meidam_2014,Glampedakis:2017dvb}

In passing we note a potential further challenge at a practical level; namely, evaluating of high derivatives
in an accurate fashion requires sufficient precision. Otherwise, a significant loss of accuracy might ensue.
This point can be relevant in deciding the most convenient discretization technique at the numerical level.
Alternatively, it is tempting to employ field redefinitions to (attempt to) reduce higher derivatives as
non-linear combinations of lower order ones (see e.g.~\cite{Solomon:2017nlh}). The extent to which this program would
be successful will depend on the particular theory being explored. Regardless, even if one could reduce all higher derivatives
to at most second order ones, one would still face mathematical obstructions 
(see e.g.~\cite{Papallo:2017qvl,Ripley:2019hxt,Bernard:2019fjb,Kovacs:2020ywu}). At a practical level this would require
and approach like the one explored in this work to control them.
Also, we find it important to stress a related point. Note the order reduced or the iterated form of the geometry equations
would naturally define slightly different foliations and care must be exercised to correctly draw contrasting lessons.

Finally, while our studies restricted to a particular theory and within the simpler setting of spherical
symmetry, the robustness and generalities of the techniques adopted gives strong backing for their use
in general scenarios. Future work will concentrate in this direction.

\begin{acknowledgments}

We would like to thank Pablo Bosch, Vitor Cardoso, Will East, Pau Figueras, Junwu Huang, Kenn Mattsson, Eric Poisson, Andrew Tolley and Huan Yang,  
for discussions during this work. 
This research was supported in part by CIFAR, NSERC through a Discovery grant, and by Perimeter Institute
for Theoretical Physics. Research at Perimeter Institute is supported by the
Government of Canada and by the Province of Ontario through the Ministry of
Research, Innovation and Science.

\end{acknowledgments}

\appendix

\section{Discrete expressions}\label{app:derivatives}

The discrete expression for second spatial derivatives satisfying SBP~\cite{MATTSSON2004503} reads, 
\setcounter{MaxMatrixCols}{12}
\begin{widetext}
 \begin{equation*}
D_{2} = \frac{1}{dx^2}
\begin{pmatrix}
\frac{114170}{40947} & -\frac{438107}{54596} & \frac{336409}{40947} & -\frac{276997}{81894} & \frac{3747}{13649} & \frac{21035}{163788} & 0 & 0  & 0 &0  &0  &\cdots \\
\phantom{a} & \phantom{a}  & \phantom{a}  & \phantom{a}  & \phantom{a}  & \phantom{a}  & \phantom{a}  & \phantom{a}   & \phantom{a}  &\phantom{a}  \\
\frac{6173}{5860} & -\frac{2066}{879} & \frac{3283}{1758} & -\frac{303}{293} & \frac{2111}{3516} & -\frac{601}{4395} &0 &0 &0 &0  &0 &\cdots  \\
\phantom{a} & \phantom{a}  & \phantom{a}  & \phantom{a}  & \phantom{a}  & \phantom{a}  & \phantom{a}  & \phantom{a}   & \phantom{a}  &\phantom{a}  \\
-\frac{52391}{81330}  & \frac{134603}{32532}  & -\frac{21982}{2711} & \frac{112915}{16266} & -\frac{46969}{16266} &  \frac{30409}{54220} &0 &0 &0 &0  &0 &\cdots\\
\phantom{a} & \phantom{a}  & \phantom{a}  & \phantom{a}  & \phantom{a}  & \phantom{a}  & \phantom{a}  & \phantom{a}   & \phantom{a}  &\phantom{a}  \\
\frac{68603}{321540}  & -\frac{12423}{10718}  & \frac{112915}{32154} & -\frac{75934}{16077} & \frac{53369}{21436} & -\frac{54899}{160770} & \frac{48}{5359} & 0 & 0 & 0  &0 &\cdots \\
\phantom{a} & \phantom{a}  & \phantom{a}  & \phantom{a}  & \phantom{a}  & \phantom{a}  & \phantom{a}  & \phantom{a}   & \phantom{a}  &\phantom{a}  \\
-\frac{7053}{39385}  & \frac{86551}{94524}  & -\frac{46969}{23631} & \frac{53369}{15754} & -\frac{87904}{23631} & \frac{820271}{472620} & -\frac{1296}{7877} &\frac{96}{7877} & 0 & 0 &0  &\cdots \\
\phantom{a} & \phantom{a}  & \phantom{a}  & \phantom{a}  & \phantom{a}  & \phantom{a}  & \phantom{a}  & \phantom{a}   & \phantom{a}  &\phantom{a}  \\
\frac{21035}{525612}  & -\frac{24641}{131403}  & \frac{30409}{87602} & -\frac{54899}{131403} & \frac{820271}{525612} & -\frac{117600}{43801} & \frac{64800}{43801} & -\frac{6480}{43801} & \frac{480}{43801} &0  &0  &\cdots \\
\phantom{a} & \phantom{a}  & \phantom{a}  & \phantom{a}  & \phantom{a}  & \phantom{a}  & \phantom{a}  & \phantom{a}   & \phantom{a}  &\phantom{a}  \\
0  & 0  & 0 & \frac{1}{90} & -\frac{3}{20} &\frac{3}{2} & -\frac{49}{18} & \frac{3}{2} & -\frac{3}{20}& \frac{1}{90} &0 & \cdots\\
0 & 0 & 0 & 0 & \ddots &  \ddots &  \ddots &  \ddots & \ddots & \ddots  &\ddots &\ddots
\end{pmatrix}
\end{equation*}
\end{widetext}
\setcounter{MaxMatrixCols}{17}
which is of 3rd order accuracy in the boundaries and 6th order in the interior.

The discrete expression for third spatial derivatives satisfying SBP~\cite{MATTSSON2014432} reads,
{\fontsize{4.999}{4}\selectfont
\begin{widetext}
\begin{equation*}
D_{3}=\frac{1}{dx^3}
\begin{pmatrix}
-\frac{151704}{63673} & \frac{15270676769}{1821047800} & -\frac{443349971}{41387450} & \frac{2063356637}{364209560} & -\frac{39300617}{45526195} & -\frac{11473393}{364209560} & -\frac{38062741}{455261950} & \frac{40315779}{1821047800} & 0 &0 & 0 &0 &0 &0  &\cdots \\
\phantom{a} & \phantom{a}  & \phantom{a}  & \phantom{a}  & \phantom{a}  & \phantom{a}  & \phantom{a}  & \phantom{a}   & \phantom{a}  &\phantom{a}  \\
-\frac{13333381409}{8182998824}&\frac{829440}{145979}&-\frac{8702160983}{1168999832}&\frac{1321219979}{292249958}&-\frac{1463113021}{1168999832}&\frac{1240729}{20874997}&\frac{102110955}{1168999832}&-\frac{50022767}{2045749706} & 0 &0 & 0 &0 &0 &0  &\cdots\\
\phantom{a} & \phantom{a}  & \phantom{a}  & \phantom{a}  & \phantom{a}  & \phantom{a}  & \phantom{a}  & \phantom{a}   & \phantom{a}  &\phantom{a}  \\
\frac{14062931}{75990642}&-\frac{1261072297}{477655464}&\frac{1088640}{139177}&-\frac{4530616889}{477655464}&\frac{602572103}{119413866}&-\frac{116503713}{159218488}&-\frac{17846623}{59706933}&\frac{343537955}{3343588248}& 0 &0 & 0 &0 &0 &0  &\cdots\\
\phantom{a} & \phantom{a}  & \phantom{a}  & \phantom{a}  & \phantom{a}  & \phantom{a}  & \phantom{a}  & \phantom{a}   & \phantom{a}  &\phantom{a}\\
\frac{661223855}{7727471752}&-\frac{214194059}{275981134}&\frac{1209539129}{1103924536}&\frac{645120}{964969}&-\frac{2321979501}{1103924536}&\frac{327603877}{275981134}&-\frac{175223717}{1103924536}&\frac{1353613}{965933969}& 0 &0 & 0 &0 &0 &0  &\cdots\\
\phantom{a} & \phantom{a}  & \phantom{a}  & \phantom{a}  & \phantom{a}  & \phantom{a}  & \phantom{a}  & \phantom{a}   & \phantom{a}  &\phantom{a}  \\
-\frac{91064195}{594070477}&\frac{632843581}{678937688}&-\frac{446896583}{169734422}&\frac{2045223021}{678937688}&\frac{22680}{593477}&-\frac{1804641793}{678937688}&\frac{311038417}{169734422}&-\frac{1932566239}{4752563816}&\frac{21168}{593477}& 0 &0 & 0 &0  &0  &\cdots\\
\phantom{a} & \phantom{a}  & \phantom{a}  & \phantom{a}  & \phantom{a}  & \phantom{a}  & \phantom{a}  & \phantom{a}   & \phantom{a}  &\phantom{a} \\.
\frac{11473393}{1249464216}&-\frac{8685103}{111559305}&\frac{116503713}{297491480}&-\frac{327603877}{223118610}&\frac{1804641793}{892474440}&0&-\frac{1760949511}{892474440}&\frac{2105883973}{1561830270}&-\frac{72576}{260045}&\frac{7056}{260045}& 0 &0 & 0 &0  &\cdots \\
\phantom{a} & \phantom{a}  & \phantom{a}  & \phantom{a}  & \phantom{a}  & \phantom{a}  & \phantom{a}  & \phantom{a}   & \phantom{a}  &\phantom{a} \\
\frac{38062741}{1420348930}&-\frac{20422191}{162325592}&\frac{17846623}{101453495}&\frac{175223717}{811627960}&-\frac{311038417}{202906990}&\frac{1760949511}{811627960}&0&-\frac{1081094773}{516490520}&\frac{1022112}{709465}&-\frac{217728}{709465}&\frac{21168}{709465}& 0 &0  &0  &\cdots \\
\phantom{a} & \phantom{a}  & \phantom{a}  & \phantom{a}  & \phantom{a}  & \phantom{a}  & \phantom{a}  & \phantom{a}   & \phantom{a}  &\phantom{a} \\
-\frac{40315779}{5832758360}&\frac{50022767}{1458189590}&-\frac{68707591}{1166551672}&-\frac{1353613}{729094795}&\frac{1932566239}{5832758360}&-\frac{2105883973}{1458189590}&\frac{1081094773}{530250760}&0&-\frac{10329984}{5098565}&\frac{7154784}{5098565}&-\frac{1524096}{5098565}&\frac{148176}{5098565} &0 &0  &\cdots\\
\phantom{a} & \phantom{a}  & \phantom{a}  & \phantom{a}  & \phantom{a}  & \phantom{a}  & \phantom{a}  & \phantom{a}   & \phantom{a}  &\phantom{a} \\
0&0&0&0&-\frac{7}{240}&\frac{3}{10}&-\frac{169}{120}&\frac{61}{30}&0&-\frac{61}{30}&\frac{169}{120}&-\frac{3}{10}&\frac{7}{240} &0  &\cdots\\
\phantom{a} & \phantom{a}  & \phantom{a}  & \phantom{a}  & \phantom{a}  & \phantom{a}  & \phantom{a}  & \phantom{a}   & \phantom{a}  &\phantom{a} \\
0 & 0 & 0 & 0 &0 & \ddots&  \ddots &  \ddots &  \ddots & \ddots & \ddots & \ddots & \ddots & \ddots & \ddots 
\end{pmatrix}
\end{equation*}
\end{widetext}

}
which is of 3rd order accuracy in the boundaries and 6th order in the interior.

\section{Convergence}
\begin{figure}[tb]
\includegraphics[width=\linewidth]{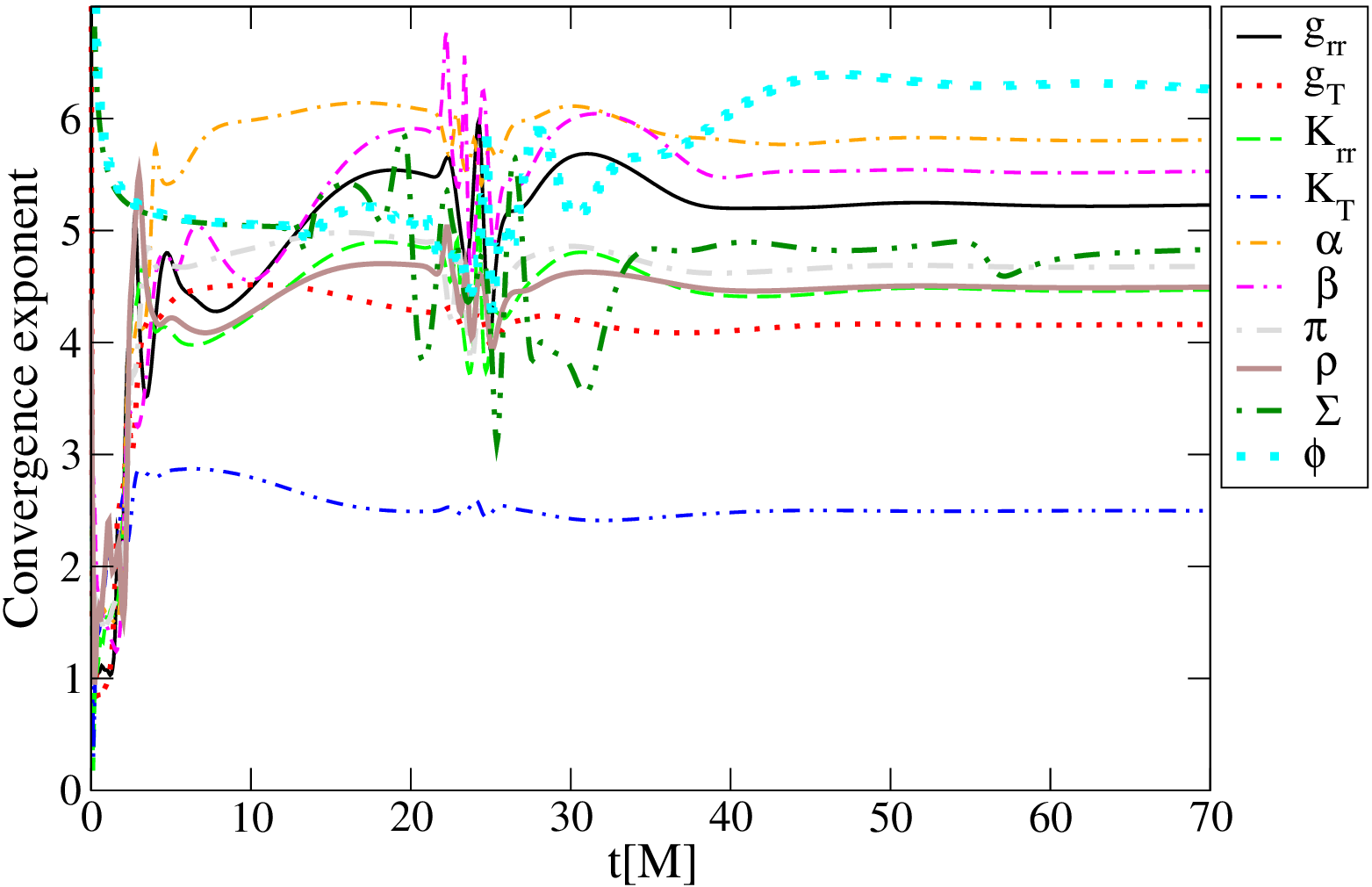}
\caption{Convergence factor $Q$ as a function of time. In most cases the convergence settles between 4th and 6th order. For $K_{T}$ $Q\approx3$.}
\label{fig:convergence}
\end{figure}

\begin{figure}[tb]
\includegraphics[width=\linewidth]{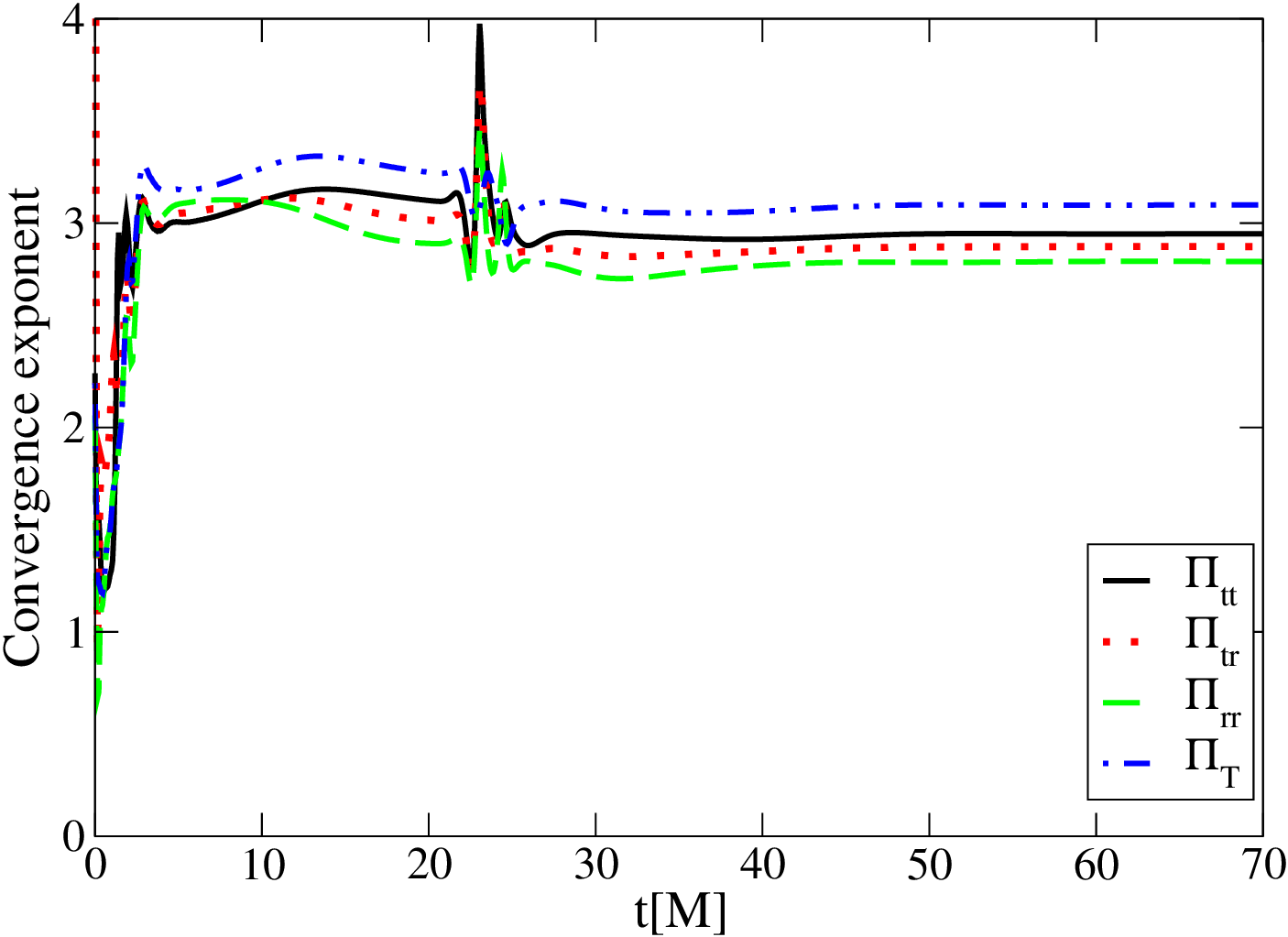}
\caption{Convergence factor $Q$ as a function of time for the $\Pi_{\mu\nu}$ variables. The behavior is consistent with 3rd order convergence.\vspace{7mm}}
\label{fig:convergence_Pi}
\end{figure}

To check convergence, we adopt the base uniform grid spacing to be $dx=0.037M$ an compute the convergence factor as,
\begin{equation}
 Q\equiv\ln{\left( \frac{||u_{dx} - u_{dx/2}||_{2}}{||u_{dx/2} - u_{dx/4}||_{2}}\right)} /\ln(2),
\end{equation}
were $u_{dx}$, $u_{dx/2}$  and $u_{dx/4}$ stands for any of the dynamical fields evolved with resolutions $dx $, $dx/2$ and $dx/4$ respectively. 
In Figures~\ref{fig:convergence} and~\ref{fig:convergence_Pi}  we present the convergence factor $Q$ for simulations with a fixed coupling of $\epsilon=1\times10^{-2}$, $\tau=5\times10^{-3}$, an initial amplitude of the scalar field given by $A=1\times10^{-3}$, 
centered at $r_{c}=20M$ and of width $\sigma=1$, and the initial total mass of the spacetime is $M=1$. Figure~\ref{fig:convergence} shows
the measured rate for the ``standard'' fields (i.e. those that would only be present in GR). The majority of fields display a
rate of around 4th to 6th order, which is consistent with the 4th order accuracy of our time integrator or the 6th order accuracy --at
interior points---of our finite difference derivative operators. The field 
$K_T$ rate is $\simeq 3$, indicating its behavior is dominated by the 3rd order accuracy at boundary points of our scheme.
Figure~\ref{fig:convergence_Pi} displays the rate for the new variables $\Pi_{\mu\nu}$ introduced to evolve the modified theory, which
converge at order $Q\approx3$.

\section{Constraints}

\begin{figure}[tb]
\includegraphics[width=\linewidth]{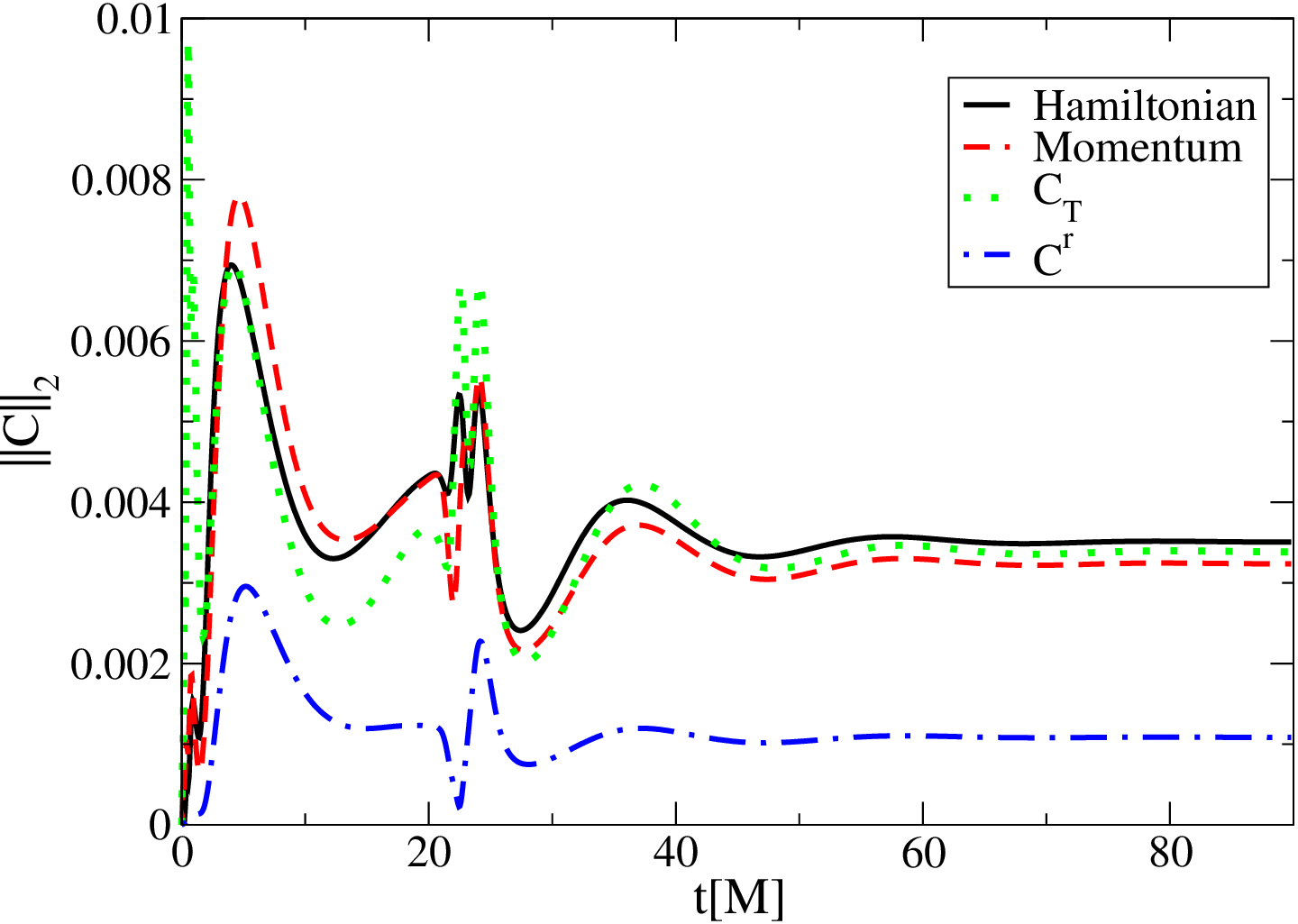}
\caption{Norm of the constraints as a function of time for $dx=0.019M$, $\epsilon=1\times10^{-2}$, $A=1\times10^{-3}$, $r_c=20M$ and $\sigma=1$.}
\label{fig:constraints}
\end{figure}

We also monitor the behavior of constraints \eqref{const1}, \eqref{const2},\eqref{const3} and \eqref{const4} during evolution. In particular Figure~\ref{fig:constraints} displays the norms of each one as a function of time 
for our base resolution of $dx=0.019M$, coupling $\epsilon=1\times10^{-2}$, $\tau=5\times10^{-3}$ and initial scalar profile with center and width are $A=1\times10^{-3}$, $r_c=20M$ and $\sigma=1$ respectively. To 
assess the magnitude of constraint violations we normalized the norms of every constraint by the sum of the norms of each term that define it. 
Such violations remain below $\approx1\%$ during evolution.

It is also important to check how effective equations \eqref{eq_rep_first_order_Pi} are to enforce variables $\Pi_{\mu\nu}$ approximate  
$\widetilde{M}_{\mu\nu}$. To this end we monitor the quantities given by,
\begin{equation}\label{control_P}
\mathcal{P}_{\mu\nu} \equiv \frac{|| \Pi_{\mu\nu} - \widetilde{M}_{\mu\nu}||_{2}}{|| \widetilde{M}_{\mu\nu}||_{2}}. 
\end{equation}

\begin{figure}[tb]
 \includegraphics[width=\linewidth]{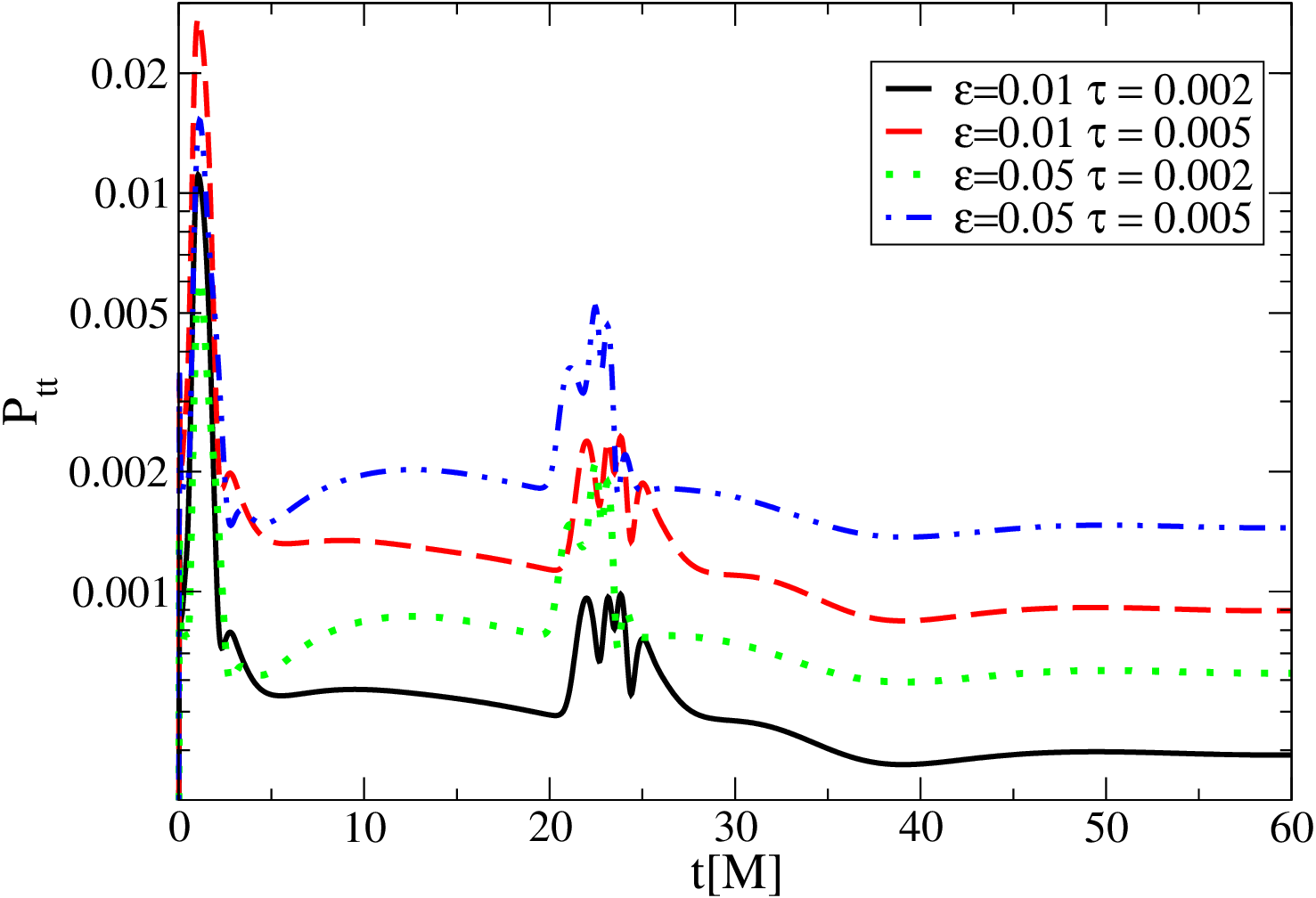}
  \caption{$\mathcal{P}_{tt}$ as a function of time for different values of the coupling $\epsilon$ and the coupling $\tau$.\vspace{5mm}}
\label{fig:control_Mtt_all}
\end{figure}

Figure~\ref{fig:control_Mtt_all} displays the behavior of $\mathcal{P}_{tt}$ for $\epsilon=\{0.01,0.05\}$ (two strong coupling values)
and choosing $\tau=\{0.002, 0.005\}$ (two different values of driving timescales). 
The difference between $\Pi_{tt}$ and $\widetilde{M}_{tt}$ stays small throughout, but it is most pronounced at two moments during the evolution. One at 
the  beginning of the simulation until the initial solution rapidly transitions (mostly due to gauge evolution) and a second rise, during the accreting stage. Both are the 
regimes with the most marked time dependence. For the values chosen, the differences are bounded by $2\%$ ($0.5\%$) during the initial (accretion) stage, but are  
diminished by decreasing the value of the timescale $\tau$.

\begin{figure}[tb]
\includegraphics[width=\linewidth]{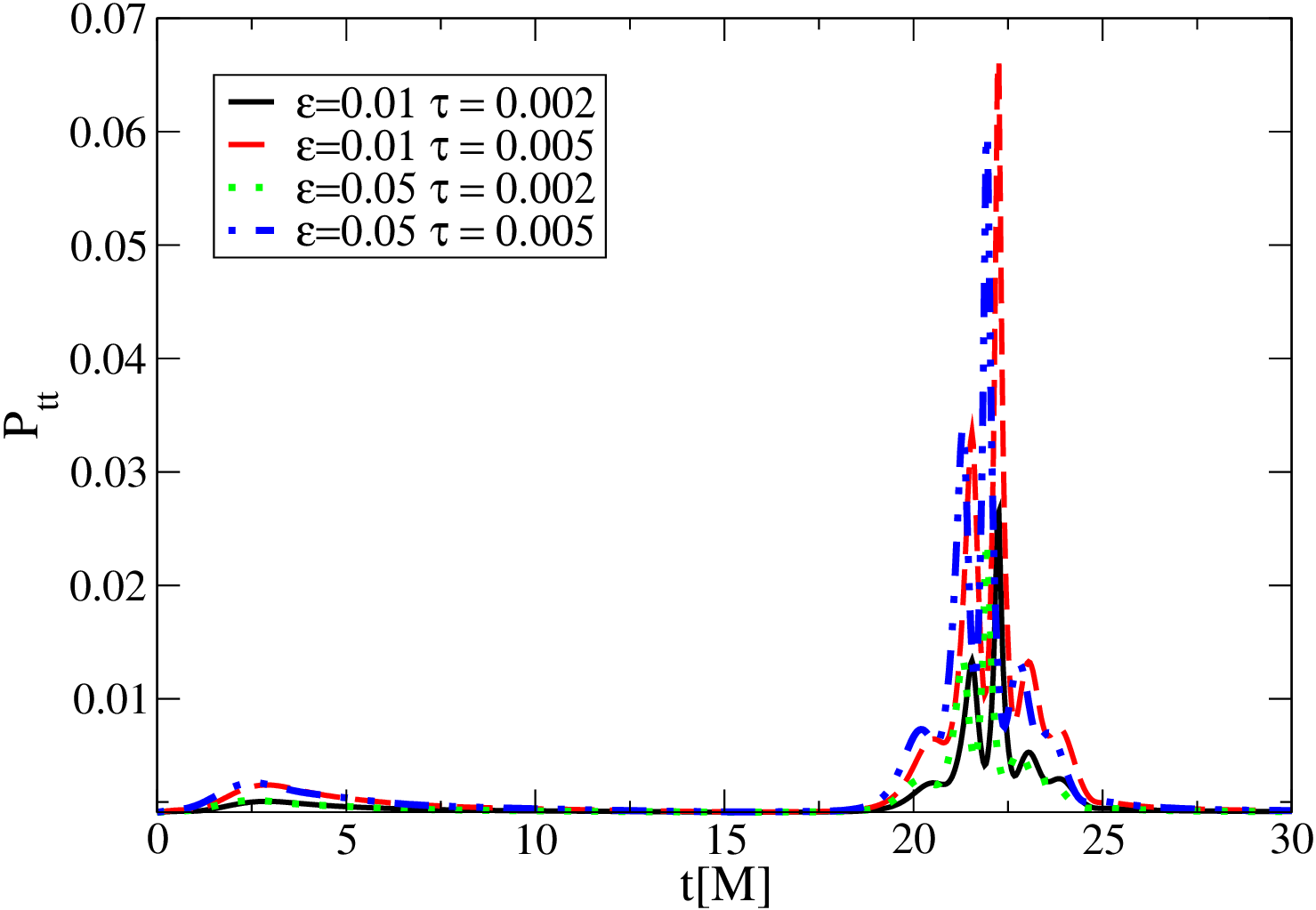}
\caption{$\mathcal{P}_{tt}$ as a function of time for different values of the coupling $\epsilon$ and the coupling $\tau$. The norms are calculated
over points at or exterior to the apparent horizon.}
\label{fig:control_Mtt_out}
\end{figure}

It is instructive also to monitor these quantities restricted to the {\em exterior} of the apparent horizon as $M_{tt}$ can be
quite large inside and skew the interpretation of difference. Figure~\ref{fig:control_Mtt_out} shows that with this restriction,
the initial transient transient is significantly reduced but it is larger during the accretion stage, raising to  $\approx7\%$. Nevertheless
this can be reduced by adopting a different value of $\tau$. For instance, it is reduced by about half going from $\tau=0.05$ to $\tau=0.02$.
Finally, we note that differences in the other components of $\mathcal{P}_{\mu\nu}$ behave similarly to the one displayed by $\mathcal{P}_{tt}$.

\clearpage
\bibliography{mybib}

\end{document}